\documentclass[superscriptaddress,prl,twocolumn,amsmath,amssymb]{revtex4-1}
\pdfoutput=1
\usepackage{graphicx}
\usepackage{eurosym}
\usepackage{amsmath}
\usepackage{amssymb}
\usepackage{dcolumn}
\usepackage{bm}
\usepackage{subfigure}

\def\bra#1{\mathinner{\langle{#1}|}}
\def\ket#1{\mathinner{|{#1}\rangle}}
\newcommand{\braket}[2]{\langle #1|#2\rangle}

\newcommand{\ignore}[1]{}

\newcommand{\be}{\begin{equation}}
\newcommand{\ee}{\end{equation}}
\newcommand{\ba}{\begin{eqnarray}}
\newcommand{\ea}{\end{eqnarray}}

\begin{document}

\title{Feynman's clock, a new variational principle, and parallel-in-time quantum dynamics}

\author{Jarrod R. McClean} 
\affiliation{Department of Chemistry and Chemical Biology, Harvard University, Cambridge MA, 02138}
\author{John A. Parkhill}
\affiliation{Department of Chemistry and Chemical Biology, Harvard University, Cambridge MA, 02138}
\author{Al\'an Aspuru-Guzik}
\affiliation{Department of Chemistry and Chemical Biology, Harvard University, Cambridge MA, 02138}

\begin{abstract} 
We introduce a new discrete-time variational principle inspired by the quantum clock originally proposed by Feynman, and use it to write down quantum evolution as a ground state eigenvalue problem. The construction allows one to apply ground state quantum many-body theory to quantum dynamics, extending the reach of many highly developed tools from this fertile research area.  Moreover this formalism naturally leads to an algorithm to parallelize quantum simulation over time. We draw an explicit connection between previously known time-dependent variational principles and the new time embedded variational principle presented. Sample calculations are presented applying the idea to a Hydrogen molecule and the spin degrees of freedom of a model inorganic compound demonstrating the parallel speedup of our method as well as its flexibility in applying ground-state methodologies.  Finally, we take advantage of the unique perspective of the new variational principle to examine the error of basis approximations in quantum dynamics.
\end{abstract}

\maketitle

\subsection{Introduction}
Feynman proposed a revolutionary solution to the problem of quantum simulation: use quantum computers to simulate quantum systems. While this strategy is powerful and elegant, universal quantum computers may not be available for some time, and in fact, accurate quantum simulations may be required for their eventual construction.  In this work, we will use the clock Hamiltonian originally introduced by Feynman\cite{Feynman:1982,Feynman:1985} for the purposes of quantum computation, to re-write the quantum dynamics problem as a ground state eigenvalue problem. We then generalize this approach, and obtain a novel variational principle for the dynamics of a quantum system and show how it allows for a natural formulation of a parallel-in-time quantum dynamics algorithm. Variational principles play a central role in the development and study of quantum dynamics\cite{Lippmann:1950ly,Heller:1976ve,Kerman:1976bh,Jackiw:1979zr,Balian:1981ys,Deumens:1994qf,Poulsen:2011kl,Haegeman:2011tg}, and the variational principle presented here extends the arsenal of available tools by allowing one to directly apply efficient approximations from the ground state quantum many-body problem to study dynamics.

Following trends in modern computing hardware, the simulations of quantum dynamics on classical hardware must be able to make effective use of parallel processing. We will show below that the perspective of the new variational principle leads naturally to a time-parallelizable formulation of quantum dynamics.
Previous approaches for recasting quantum dynamics as a time-independent problem include Floquet theory for periodic potentials\cite{Milfeld:1983,Autler:1955,Dion:2007} and more generally the $(t, t')$ formalism of Peskin and Moiseyev\cite{Peskin:1993}. However, the approach proposed in this manuscript differs considerably from these previous approaches. We derive our result from a different variational principle, and in our embedding the dynamics of the problem are encoded directly in its solution, as opposed to requiring the construction of another propagator. Considerable work has now been done in the migration of knowledge from classical computing to quantum computing and quantum information\cite{Nielsen:2000,Aspuru:2005,Kassal:2008,Wang:2008,Kassal:2009}.  In this paper, we propose a novel use of an idea from quantum computation for the simulation of quantum dynamics.  

The paper is organized as follows. We will first review the Feynman clock: a mapping stemming from the field of quantum computation that can be employed for converting problems in quantum evolution into ground-state problems in a larger Hilbert space.  We then generalize the Feynman clock into a time-embedded discrete variational principle (TEDVP) which offers additional insight to quantum time-dynamics in a way that is complementary to existing differential variational principles \cite{Dirac:1930,Frenkel:1934,McLachlan:1964}. We then apply the configuration interaction method \cite{Slater:1929,Boys:1950} from quantum chemistry to solve for approximate dynamics of a model spin system demonstrating convergence of accuracy of our proposed approach with level of the truncation. We demonstrate how this construction naturally leads to an algorithm that takes advantage of parallel processing in time, and show that it performs favorably against existing algorithms for this problem. 
Finally we discuss metrics inspired by our approach that can be used to quantitatively understand the errors resulting from truncating the Hilbert space of many-body quantum dynamics.

\subsection{Physical dynamics as a sequence of quantum gates}
Consider a quantum system described by a time-dependent wavefunction $\ket{\Psi(t)}$.
The dynamics of this system are determined by a Hermitian Hamiltonian $H(t)$ according
to the time-dependent Schr\"odinger equation in atomic units,
\begin{equation}
 i \partial_t \ket{ \Psi(t)} = H(t) \ket{\Psi(t)} 
\end{equation}
A formal solution to the above equation can be generally written: 
\begin{equation}
 \ket{\Psi(t)} = \mathcal{T} \left( e^{-i \int_{t_0}^t dt' H(t')} \right) \ket{\Psi(t_0)} = U(t,t_0) \ket{\Psi(t_0)} 
\end{equation}
Where $\mathcal{T}$ is the well known time-ordering operator and $U(t,t_0)$ is a 
unitary operator that evolves the system from a time $t_0$ to a
time $t$.  These operators also obey a group composition property, such that if 
$t_0 < t_1 < ... < t_n < t$ then
\begin{align}
 \ket{\Psi(t)} = U(t,t_0) \ket{\Psi(t_0)} = \notag \\
 U(t,t_n) U(t_n, t_{n-1}) ... U(t_1, t_0) \ket{\Psi(t_0)}
\end{align}
From the unitarity of these operators, it is of course also true that 
$U(t_n, t_{n-1})^\dagger = U(t_{n-1}, t_n)$ where $\dagger$ indicates the adjoint.
Thus far, we have treated time as a continuous variable.  However, when one considers numerical
calculations on a wavefunction, it is typically necessary to discretize time.  

We discretize time by keeping an ancillary quantum system, which can occupy states with integer indices ranging
from $\ket{0}$ to $\ket{T-1}$ where $T$ is the number of discrete time steps under consideration.
This quantum system has orthonormal states such that
\begin{equation}
 \braket{i}{j} = \delta_{ij}
\end{equation} 
This definition allows one to encode the entire evolution of a physical system by entangling the
physical wavefunction with this auxiliary quantum system representing time, known as the ``time register''. We define this compound state to be the history state, given by
\begin{equation}
 \ket{\Phi} = \frac{1}{\sqrt{T}} \sum_t \ket{\Psi_t} \otimes \ket{t}
\end{equation}
where subscripts are used to emphasize when we are considering a time-independent
state of a system at time $t$. That is, we define $\ket{\Psi_{i}} = \ket{\Psi(t)}|_{t=t_i}$.
From these definitions, it is immediately clear from above that the wavefunction at any time $t$ can
be recovered by projection with the time register, such that
\begin{equation}
 \ket{\Psi(t)}|_{t=t_i} = \sqrt{T}\braket{i}{\Phi}
\end{equation}
Additionally, we discretize the action of our unitary operators, such that $U(t_1, t_0) = U_0$
and we embed this operator into the composite system-time Hilbert space as 
$\left(U_0 \otimes \ket{1} \bra{0} \right)$.  While the utility of this discretization has not yet been made apparent, we will now use this discretization to 
transform the quantum dynamics problem into a ground state eigenvalue problem.

\subsection{Feynman's clock}
In the gate model\cite{Nielsen:2000,Farhi:2000} of quantum computation, one begins with
an initial quantum state $\ket{\Psi_0}$ state, applies a sequence of unitary operators
$\{U_i\}$, known as quantum gates. By making a measurement on the final state $\ket{\Psi_f}$, one determines the result of the computation, or equivalently the result of applying the sequence of unitary operators $\{U_i\}$. The map from the sequence of unitary operators $\{U_i\}$ in the gate model, to a Hamiltonian that has the clock state as its lowest eigenvector is given by a construction called the Clock Hamiltonian\cite{Feynman:1985}.  Since its initial inception,
much work has been done on the specific form of the clock, making it amenable to implementation
on quantum computers\cite{Kitaev:2002}. However, for the purposes of our discussion that pertain to implementation
on a classical computer, the following simple construction suffices
\begin{align}
  \mathcal{H} = C_0 + \frac{1}{2} \sum_{t=0}\left( I \otimes \ket{t} \bra{t} - U_t \otimes \ket{t+1} \bra{t} \right. \notag \\
   \left. - U_t^\dagger \otimes \ket{t} \bra{t+1} + I \otimes \ket{t+1} \bra{t+1} \right)
\end{align}
where $C_0$ is a penalty term which can be used to specify the state of the physical system at any time.
Typically, we use this to enforce the initial state, such that if the known state at time $t=0$ is given
by $\ket{\Psi_0}$, then
\begin{equation}
  C_0 = (I - \ket{\Psi_0}\bra{\Psi_0}) \otimes \ket{0} \bra{0}
\end{equation}

One may verify by action of $\mathcal{H}$ on the history state defined above, $\ket{\Phi}$, that the history
state is an eigenvector of this operator, with eigenvalue $0$.  

\begin{figure}
\includegraphics[width=8.0 cm]{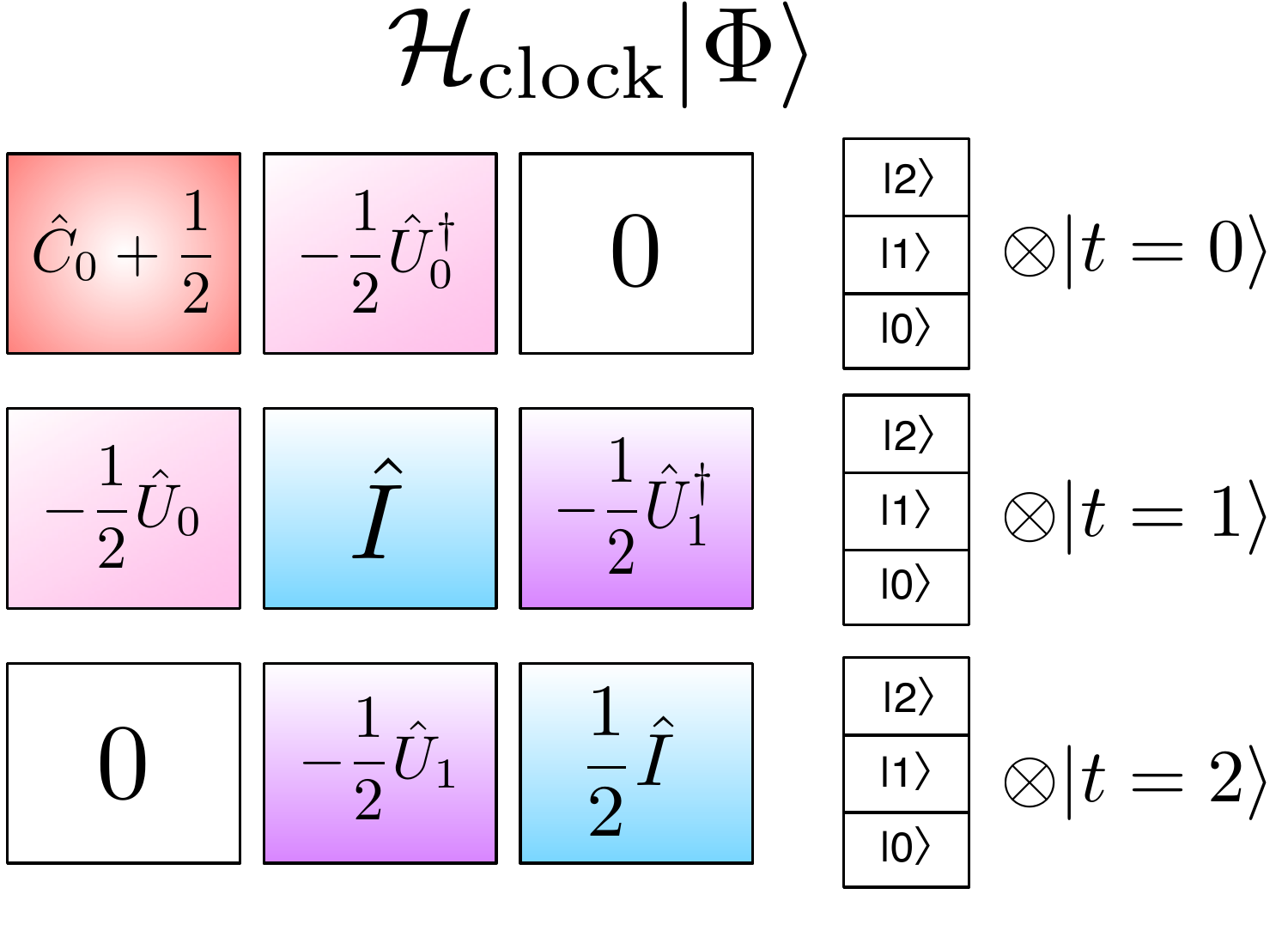}
\caption{A schematic representation of the action of the clock Hamiltonian on the history state with three discrete times, and a Hilbert space of three states. Each block is a matrix with dimension of the physical system.}
\label{fig:hclock}
\end{figure}

\subsection{A discrete-time variational principle}
We now introduce a new time-embedded discrete variational principle (TEDVP) and show how the above eigenvalue problem emerges as the special case of choosing a linear variational space. 
Suppose that one knows the exact form of the evolution operators and wavefunctions at times 
$t=0,1$.  By the properties of unitary evolution it is clear that the following holds:
\begin{equation}
  2 - \bra{\Psi_1} U_0 \ket{\Psi_0} - \bra{\Psi_0} U_0^\dagger \ket{\Psi_1} = 0
\end{equation}

From this, we can see that if the exact construction of $U_i$ is known for all $i$, but the wavefunctions
are only approximately known(but still normalized), it is true that
\begin{equation}
  \sum_{t=0}^{T-1} \left(2 - \bra{\Psi_{t+1}} U_t \ket{\Psi_t} - \bra{\Psi_t} U_t^\dagger \ket{\Psi_{t+1}}\right) \geq 0
\end{equation}
where equality holds in the case that the wavefunction becomes exact at each discrete time point.
Reintroducing the ancillary time-register, we may equivalently say that all valid time evolutions
embedded into the composite system-time space as $\sum_t \ket{\Psi_t}\ket{t}$ minimize the quantity
\begin{equation}
  \mathcal{S} = \sum_{t,t'} \bra{t'}\bra{\Psi_{t'}} \mathcal{H} \ket{\Psi_t} \ket{t}
\end{equation}
where $\mathcal{H}$ (script font for operators denotes they act in the composite system-time space) is the operator given by
\begin{equation}
 \mathcal{H} = \sum_{t=0} I \otimes \ket{t} \bra{t} - U_t \otimes \ket{t+1} \bra{t}
   - U_t^\dagger \otimes \ket{t} \bra{t+1} + I \otimes \ket{t+1} \bra{t+1}
\end{equation}
It is then clear from the usual ground state variational principle, that the expectation value of the operator
\begin{equation}
  \mathcal{S} = \sum_{t,t'} \bra{t'}\bra{\Psi_{t'}} \mathcal{H} \ket{\Psi_t} \ket{t}
\end{equation}
is only minimized for exact evolutions of the physical system. This leads us immediately to a time-dependent variational principle for the discrete representation of a wavefunction given by:
\begin{equation}
 \label{DiscreteAction}
 \delta S = \delta \sum_{t,t'} \bra{t'}\bra{\Psi_{t'}} \mathcal{H} \ket{\Psi_t} \ket{t} = 0
\end{equation}It is interesting to note, that this is a true
variational principle in the sense that an exact quantum evolution is found at a minimum, rather than a stationary point as in some variational principles\cite{McLachlan:1964}.  This is related to the connection between this variational principle and the McLachlan variational principle that is explored in the next section. However, to the authors knowledge, this connection has never been explicitly made until now. 

To complete the connection to the clock mapping given above, we first note that this operator is Hermitian by construction and choose a linear variational space that spans the entire physical domain. To constrain the solution to have unit norm, we introduce the Lagrange multiplier $\lambda$ and minimize the auxiliary functional given by
\begin{equation}
  \mathcal{L} = \sum_{t,t'} \bra{t'}\bra{\Psi_{t'}} \mathcal{H} + C_0 \ket{\Psi_t} \ket{t}  - \lambda \left( \sum_{t,t'} \bra{t'}\braket{\Psi_{t'}}{\Psi_t}\ket{t} - 1 \right)
\end{equation}
It is clear that this problem is equivalent to the exact eigenvalue problem on $\mathcal{H}$ with eigenvalue $\lambda$. The optimal trajectory is given by the ground state eigenvector of the operator $\mathcal{H}$.  From this construction, we see that the clock mapping originally proposed by Feynman is easily recovered as the optimal variation of the TEDVP in a complete linear basis, under the constraint of unit norm.  Note that the inclusion of $C_0$
as a penalty term to break the degeneracy of the ground state is only a convenient construction for
the eigenvalue problem.  In the general TEDVP, one need not include this penalty explicitly,
as degenerate allowed paths are excluded, as in other time-dependent variational principles, by fixing the initial state.

We note, as in the case of the time-independent variational principle and differential formulations of the time-dependent variational principle, the most compact solutions may be derived from variational spaces that have non-linear parameterizations.  Key examples of this in chemistry include Hartree-Fock, coupled cluster theory\cite{Coester:1960,Cizek:1966,Bartlett:2007}, and multi-configurational time-dependent Hartree\cite{Beck20001}.  It is here that the generality of this new variational principle allows one to explore new solutions to the dynamics of the path without the restriction of writing the problem as an eigenvalue problem, as in the original clock construction of Feynman.

\begin{figure}
\centering
\includegraphics[width=8cm]{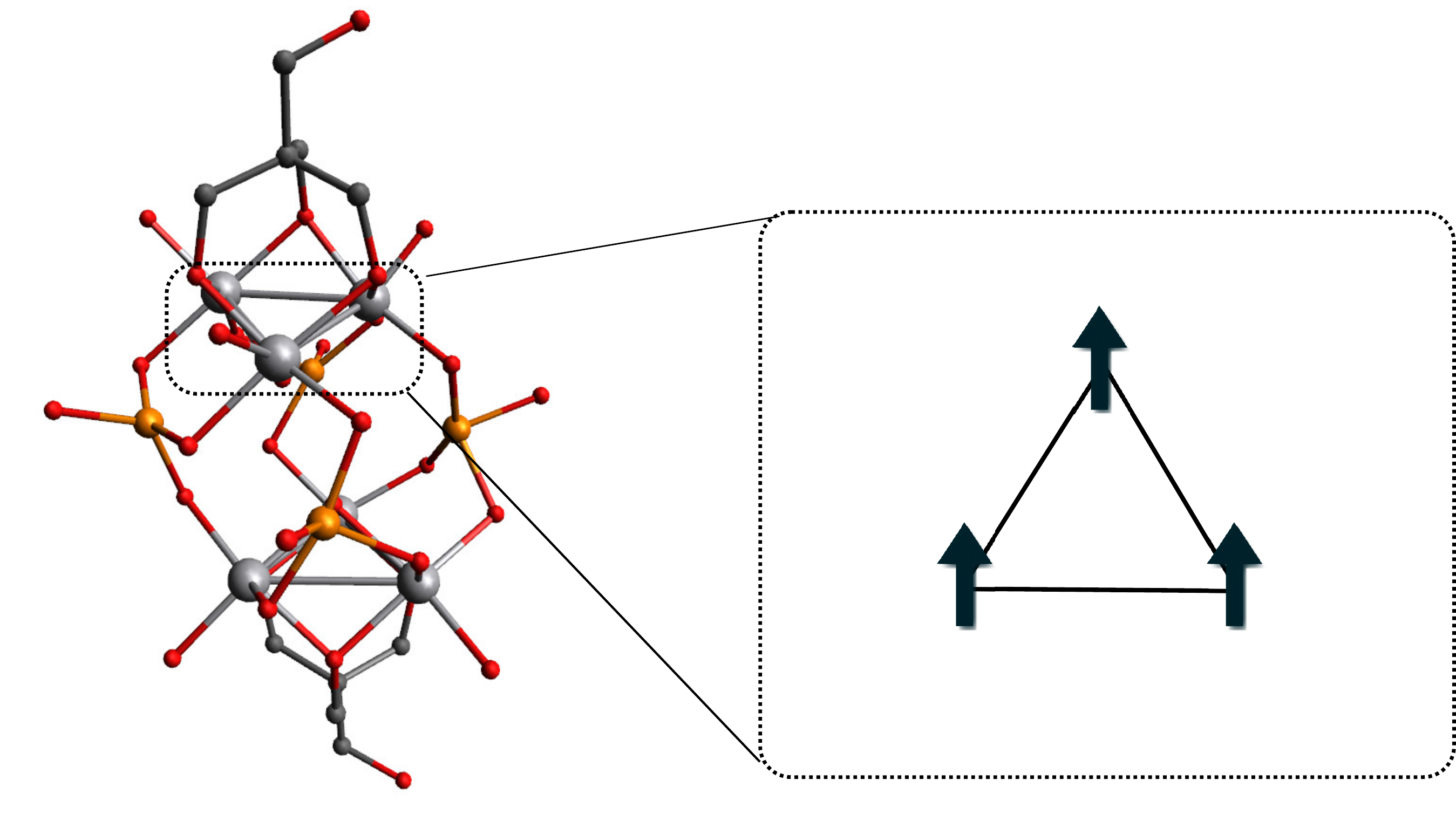}
\caption{The spin triangle within the vanadium compound, used as a model system for the TEDVP.  Note that coordinating sodium ions and water molecules are not depicted here.  The chemical formula of this compound is given by (CN$_3$H$_6)_4$Na$_2$[H$_4$V$_6^{(IV)}$O$_8$(PO$_4$)$_4$ {(OCH$_2$)$_3$ CCH$_2$OH}$_2$] $\cdot$ 14H$_2$O}
\label{fig:vanad}
\end{figure}

\subsection{Connection to previous time-dependent variational principles}
In the limit of an exact solution, it must be true that all valid time-dependent
variational principles are satisfied. For that reason, it is important to draw the formal connection between our variational principle and previously known variational principles.

Considering only two adjacent times $t$ and $t+1$, the operator $\mathcal{H}$ is given by:
\begin{align}
\label{TwoTimeClock}
 \mathcal{H} = \frac{1}{2} ( I \otimes \ket{t}\bra{t} - U_t \otimes \ket{t+1} \bra{t} \notag \\
              - U_t^\dagger \otimes \ket{t} \bra{t+1} + I \otimes \ket{t+1}\bra{t+1} )
\end{align}
Now suppose that the separation of physical times between discrete step $t$ and $t+1$, denoted
$dt$ is small, and the physical system has an associated Hamiltonian given by $H$, such that
\begin{equation}
 U_t = e^{-i H dt} \approx I - i H dt - \frac{H^2 dt^2}{2}
\end{equation}


By inserting this propagator into equation ~\ref{DiscreteAction}, rewriting the result in terms of
$\ket{\Psi(t)}$, and dropping terms of order $dt^3$ we have
\begin{align}
 &\delta \left( \braket{\Psi(t)}{\Psi(t)} - \bra{\Psi(t+dt)} \left(I - iHdt \right) \ket{\Psi(t)} \right. \notag \\
  & - \bra{\Psi(t)}\left( I + iHdt \right)\ket{\Psi(t+dt)} + \notag \\
  & \bra{\Psi(t)} H^2dt^2\ket{\Psi(t)} +  \braket{\Psi(t+dt)}{\Psi(t+dt)} )= 0
\end{align}
After defining the difference operator such that
$\partial_t \ket{\Psi(t)} \equiv \left[\ket{\Psi(t + dt)} - \ket{\Psi(t)}\right]/{dt} $,
we can factorize the above expression into
\begin{align}
 \delta \bra{\Psi(t)} \left(i \partial_t - H \right)^\dagger \left(i \partial_t - H \right) \ket{\Psi(t)} = 0
\end{align}
In the limit that $dt \rightarrow 0$, these difference operators become derivatives.
Defining $\Theta = i \partial_t \ket{\Psi(t)}$, this is precisely the McLachlan variational
principle\cite{McLachlan:1964}.  
\begin{equation}
 \delta \|\Theta - H\ket{\Psi(t)}\|^2 = 0
\end{equation}
We then conclude that in the limit of infinitesimal physical time for a single time step, the
TEDVP is equivalent to the McLachlan variational principle.  Under the reasonable assumptions that the
variational spaces of the wavefunction and its time derivative are the same and that the 
parameters are complex analytic, then it is also equivalent to the Dirac-Frenkel and Lagrange
variational principles\cite{Broeckhove:1988}.  Moreover, as supplementary material (SI1), we provide
a connection that allows other variational principles to be written as eigenvalue problems, and further discuss the merits of the integrated formalism used here.

To conclude this section, we highlight one additional difference between practical uses of the TEDVP and other variational principles: The TEDVP is independent of the
method used to construct the operator $U_t$.  In quantum information applications,
this implies it is not required to know a set of generating Hamiltonians for quantum gates.  Additionally, in numerical applications, one is not restricted by the choice of approximate propagator used. In cases where an analytic propagator is known for the chosen basis, it can be sampled explicitly.Suppose that the dimension of the physical system is given by $N$ and the number of timesteps of interest is given by $T$.  Assuming that the time register $\ket{T}$ is ordered, the resulting eigenvalue problem is block tridiagonal with dimension $NT$ (See Fig. 1).  This structure has been described at least once before in the context of ground-state quantum computation\cite{Mizel:2004}, but to the authors knowledge, never in the context of conventional simulation of quantum systems.  

\section{Many-Body application of the TEDVP}
There has been a recent rise in the interest of methods for simulating quantum spin dynamics in chemistry \cite{Kuprov:2007,Hogben:2010}. To study the properties of the clock mapping when used to formulate approximate dynamics, we chose a simple model spin system inside an inorganic molecule \cite{Luban:2002}.  Specifically, we examine the spin dynamics of the vanadium compound depicted in Fig. 2. By choosing the three unpaired electron spins to interact with one another by means of isotropic exchange as well as uniform static external magnetic field $B_0$, and a time-dependent transverse field $B_1$, this system can be modeled with a spin Hamiltonian
\begin{align}
  H = J_a (S_1 \cdot S_2 + S_1 \cdot S_3) + J_c S_2 \cdot S_3 + \notag \\
   \mu B_0 (S_{1}^z + S_{2}^z + S_{3}^z) 
  	+ \mu B_1 (S_{1}^x + S_{2}^x + S_{3}^x)
\end{align}
Where $S_{i}^\alpha$ is the $\alpha$ Pauli operator on spin $i$, $\mu = g \mu_b$, $\mu_b$ is the Bohr magneton, $g$ is the measured spectroscopic splitting factor. The couplings \( J_a \neq J_c \) as well as $g$ are fitted through experimental determinations of magnetic susceptibility\cite{Luban:2002}.  The fact that they are not equal is reflective of the isosceles geometry of the vanadium centers. The parameters of this model, are given by: $g = 1.95$, $J_a = 64.6 \pm 0.5 K$, and $J_c = 6.9 \pm 1 K$.
We will allow $B_0$ to vary to study the properties of the clock mapping in the 
solution of approximate quantum dynamics.\\

\indent 	The quantum chemistry community has decades of experience in developing and employing methods for obtaining approximate solutions of high-dimensional, ground-state eigenvector problems. By utilizing the connection we have made from dynamics to ground state problem, we will now borrow and apply the most conceptually simple approximation from quantum chemistry: configuration interaction in the space of trajectories\cite{Habershon:2012nx}, to our model problem to elucidate the properties of the clock mapping.

\begin{figure*}
\centering
\includegraphics[width = 18cm]{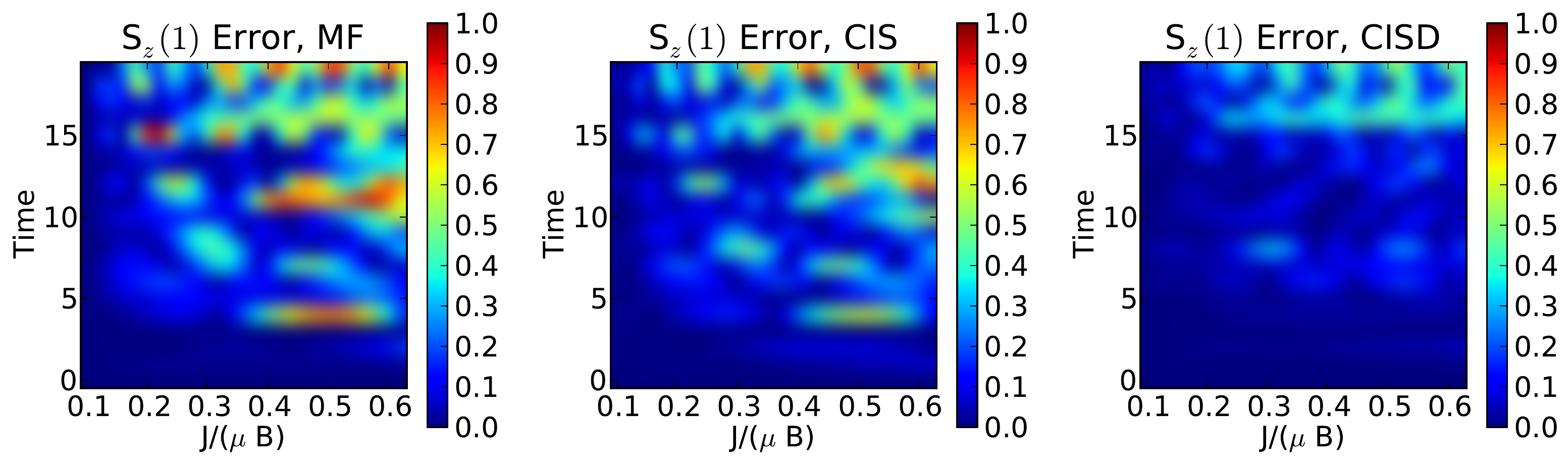}
\caption{The convergence of the CI expansion \emph{in time} is robust to strength of perturbation, and choice of initial state for the Vanadium complex. In these plots the error of the expectation value of S$_z$ at site 1 is plotted as a function of time, and coupling constant, each propagation with a different coupling constant is begun from a different random initial product state.  The expansion approaching Full CI is given by Mean Field (MF), Configuration Interaction with Single Excitations (CIS), Configuration Interaction with Single and Double Excitations(CISD), and FCI (Exact) to which the solutions are compared.}
\label{fig:errorsurface}
\end{figure*}

For the uncorrelated reference, we take the entire path of a mean-field spin evolution that is governed by the time-dependent Hartree
equations, and write it as:
\begin{equation}
  \ket{\Psi_{MF}} = \sum_t \left(\prod_i U^{i}_t \ket{0}_i\right)\ket{t} 
  	= \sum_t \left(\prod_i \ket{0}^t_i\right)\ket{t} = \sum_t \ket{\phi_t} \ket{t}
\end{equation}
where $\ket{0}_i$ is the reference spin-down state for spin $i$, $\ket{0}_i^t$ is the reference spin-down state after rotation at time $t$, $\ket{\phi_t}$ is the whole product system at
time $t$, and $U_t^{i}$ is determined from the mean field Hamiltonian. The equations of motion that determine $U_t^i$ are derived in a manner analogous to the time-dependent Hartree equations, and if one
writes the wavefunction in the physical space as

\begin{equation}
 \ket{\psi(t)} = a(t) \prod_i U_i(t) \ket{0}_i = a(t) \prod_i \ket{\phi_i} = a(t) \ket{\Phi(t)}
\end{equation}

Then the equations of motion are given by
\begin{align}
 &a(t) = a(0) \\
 &i \dot U_i = (H^{(i)} - \left( \frac{f-1}{f} \right) E(t)) U_i
\end{align}
Where $H^{(i)}$ is the mean field Hamiltonian for spin $i$ formed by contracting the Hamiltonian over
all other spins $j \neq i$, $E(t)$ is the expectation value of the Hamiltonian at time $t$, and $f$ is the
number of spins in the system. 

To generate configurations, we also introduce the transformed spin excitation operator $\tilde S_{it}^+$,
which is defined by
\begin{align}
  S_i^+ \ket{0}_i &= \ket{1}_i \\
  U^{i\dagger}_t S_i^+ U^{i\dagger}_t &= \tilde S_{it}^+ \\
  \tilde S_{it}^+ \ket{0}^t_i &= \ket{1}^t_i
\end{align}
In analogy to traditional configuration interaction, we will define different levels of excitation
(e.g. singles, doubles, ...) as the number of spin excitations at each time $t$ that will be included
in the basis in which the problem is diagonalized.  For example, the basis for the configuration
interaction singles(CIS) problem is defined as
\begin{equation}
  \mathcal{B_{CIS}} = \left\{ \tilde S_{jt}^+ \ket{\phi_t} \ket{t} 
    |\ j \in [0,n], t \in [0, T)  \right\}
\end{equation}
Note that $\tilde S_{jt}^+$ for $j=n$ is simply defined to be the identity operator on all spins
so that the reference configuration is naturally included.  Similarly, the basis for single
and double excitations(CISD) is given by

\begin{equation}
\mathcal{B_{CISD}} = \left\{ \tilde S_{jt}^+ \tilde S_{kt}^+ \ket{\phi_t} \ket{t} 
    |\ j \in [0,n], k \in [0, j), t \in [0, T)  \right\}
\end{equation}
Higher levels of excitation follow naturally, and it is clear that when one reaches a level
of excitation equivalent to the number of spins, this method may be regarded as
full configuration interaction, or exact diagonalization in the space of discretized paths.

The choice of a time-dependent reference allows the reference configuration to be nearly exact
when $B_0, B_1 >> J_a, J_c$, independent of the nature of the time-dependent transverse field.  This allows for the separation of the consequences of time-dependence from the effects of two versus one body spin interactions.

%
%

\begin{figure}
\centering
\includegraphics[width=8cm]{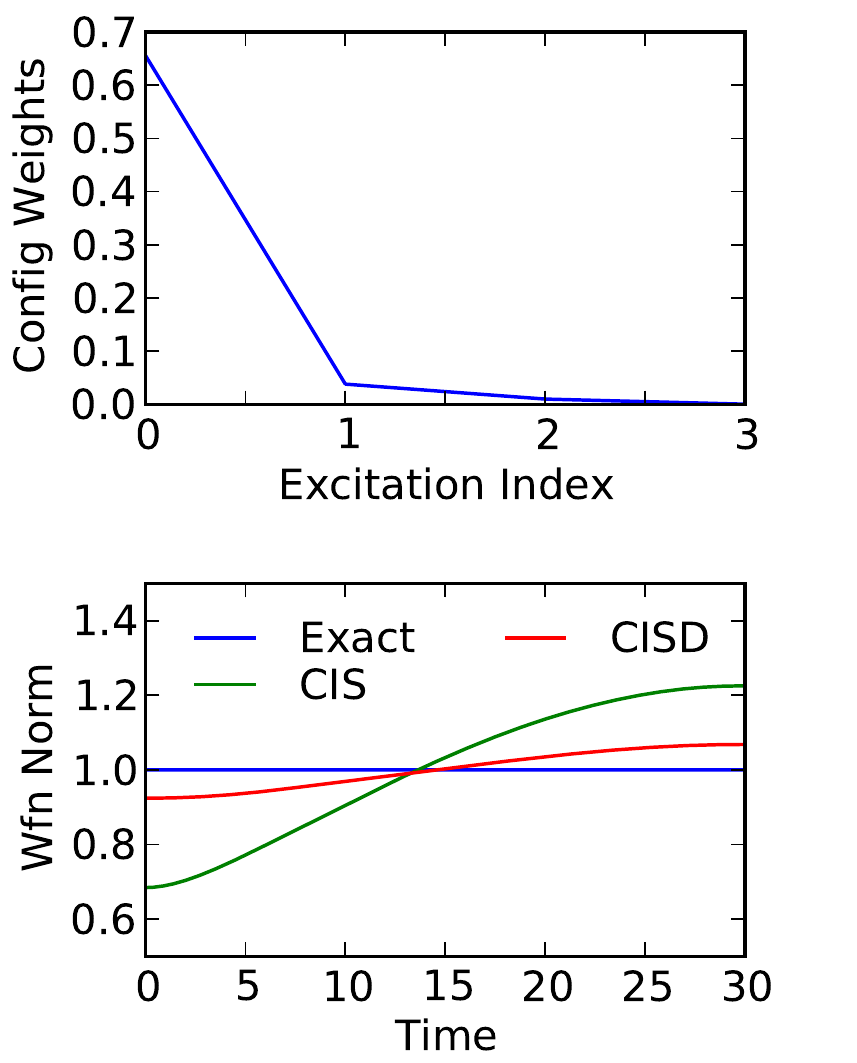}
\caption{Following a simulation of the dynamics of the Vanadium spin complex, the total contribution to the ground state eigenvector from
each level of excitation is plotted, where 0=MF(Mean-Field), 1=CIS(Configuration Interaction with Single Excitations), 2=CISD(Configuration interaction with Single and Double Excitations), and 3=FCI(Exact Diagonalization), and seen to decrease with excitation supporting the quality of the time-dependent reference state.  The units of time are $K^{-1}$.  The deviation of the wavefunction norm from unity resulting from projection is seen to decrease monotonically with level of excitation.}
\label{fig:NormConfig}
\end{figure}

After a choice of orthonormal basis, the dynamics of the physical system are given by the ground state eigenvector of the projected eigenvalue problem
\begin{equation}
  \mathcal{H}_{\mathcal{B}_i} \ket{\tilde \Phi} = E \ket{\tilde \Phi} 
\end{equation}
where we explicitly define the projection operator
onto the basis ${\mathcal{B}_i}$ as $P_{\mathcal{B}_i} = \sum_{\ket{j} \in \mathcal{B}_i} \ket{j} \bra{j}$
so that the projected operator is given by 
$\mathcal{H}_{\mathcal{B}_i} = \left(P_{\mathcal{B}_i} \mathcal{H} P_{\mathcal{B}_i}\right)$

Using these constructions, we calculate the quantum dynamics of the sample system described above.  For convenience, we rescale the Hamiltonian by the value of $1/\mu B_0$.  To add arbitrary non-trivial time dependence to the system and mimic the interaction of the system with a transverse pulse, we take $B_1 \propto \exp{(-t^2/2)} \cos (mt)$. The
magnitude of $B_0$ was taken to be $200 T$ in order to model perturbative two body interactions
in this Hamiltonian.  To propagate the equations of motion and generate the propagators for the clock operator we use the enforced time-reversal symmetry exponential propagator\cite{Castro:2004} given by
\begin{equation}
U_t = \exp \left( -i \frac{dt}{2} H(t + dt) \right) \exp \left( -i \frac{dt}{2} H(t) \right)
\end{equation}

The dynamics of some physical observables are displayed (Fig. 5) for the reference configuration, single excitations, double excitations, and full configuration interaction. The physical observables have been calculated with normalization at each time step.  It is seen (Fig. 3) that as in the case of ground state electronic structure the physical observables become more accurate both qualitatively and quantitatively as the configuration space expands, converging to the exact solution with full configuration interaction.  Moreover in Fig. 4 we plot the contributions from the reference configuration, singles space, doubles space, and triples space and observe rapidly diminishing contributions.  This suggests that the time-dependent path reference used here provides an good qualitative description of the system.  As a result, perturbative and other dynamically correlated methods from quantum chemistry may also be amenable to the solution of this problem.

\begin{figure}
\centering
\includegraphics[width=8cm]{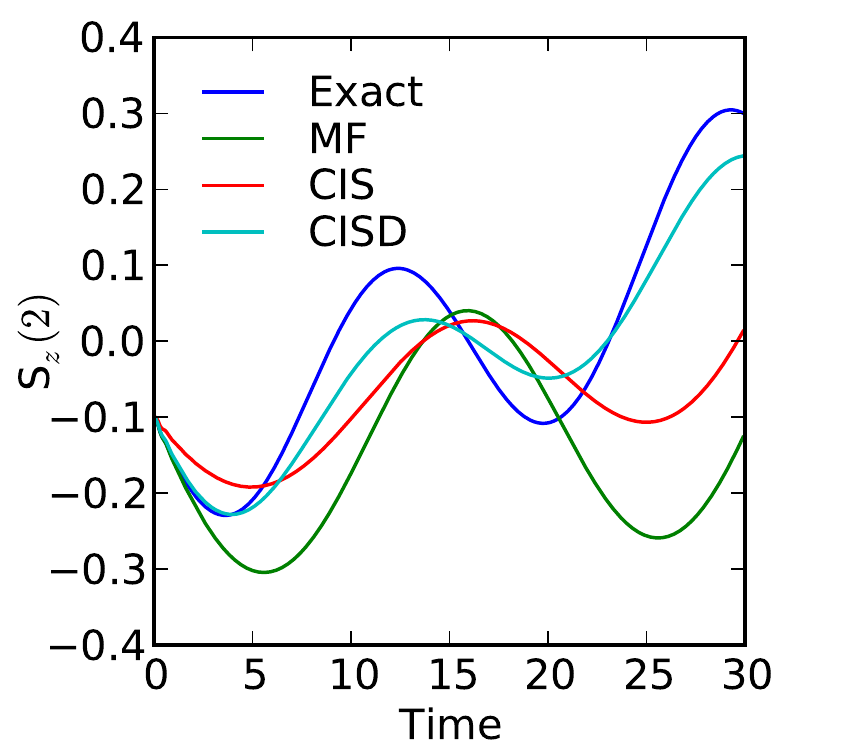}
\caption{Trajectories for single-particle observables in different levels of basis set truncation starting from the reference configuration path (MF) for the Vanadium spin complex.  The units of time are $K^{-1}$. As more levels of spin excitation are included with configuration interaction singles(CIS) and configuration interaction with singles and doubles(CISD) the trajectories converge to the exact result.}
\label{fig:OneParticle}
\end{figure}

In principle, approximate dynamics derived from this variational principle are not norm conserving, as is seen in Fig. 4, however this actually offers an important insight into a major source of error in quantum dynamics simulations of many-body systems, which is the truncation of the basis set as described in the last section. Simulations based on conventional variational principles that propagate within an incomplete configuration space easily preserve norm; however the trajectories of probability which should have left the simulated space are necessarily in error. 

\begin{figure}
\centering
\includegraphics[width=8cm]{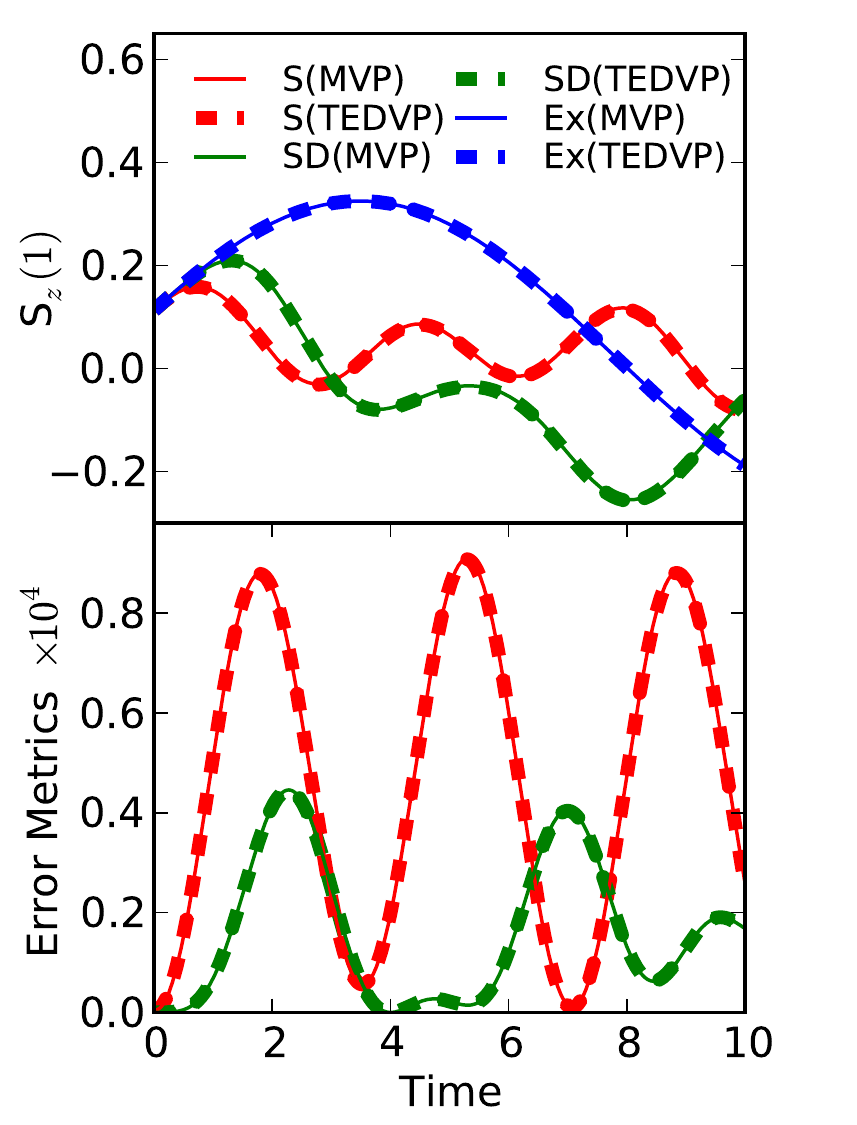}
\caption{The dynamics of a spin observable in the Vanadium spin complex at a short time step, $dt=0.01 K^{-1}$, are indistinguishable when generated with equations of motion determined by the time embedded discrete variational principle(TEDVP) (dashed lines) and the McLachlan variational principle(MVP) (solid lines).  Results are shown for propagations restricted to the space of single excitations from the initial state(S), double excitations (SD), and the full space(Ex).  The associated error metrics, $\mathcal{N}_1(t)$ for the TEDVP (dashed) and $\mathcal{N}_2(t)$ for the MVP (solid), also yield nearly identical results, displaying peaks correlated with qualitative deviations from the exact trajectory.}       
\label{fig:NormLoss}
\end{figure}

\section{Parallel-in-time quantum dynamics}
Algorithms that divide a problem up in the time domain, as opposed to spatial domain, are known as parallel-in-time algorithms.  Compared to their spatial counterparts, such as traditional domain decomposition\cite{Smith:2004}, these algorithms have received relatively little attention. This is perhaps due to the counterintuitive notion of solving for future times in parallel with the present.  However as modern computational architectures continue to evolve towards many-core setups, exploiting all avenues of parallel speedup available will be an issue of increasing importance. The most popular parallel-in-time algorithm in common use today is likely the \emph{parareal} algorithm\cite{Lions:2001,Baffico:2002}. The essential ingredients of parareal are the use of a coarse propagator $U^c$ that performs an approximate time evolution in serial, and a fine propagator $U^f$ that refines the answer and may be applied in parallel. The two propagations are combined with a predictor-corrector scheme.  It has been shown to be successful with parabolic type equations\cite{Gander:2007}, such as the heat equation, but it has found limited success with wave-like equations\cite{Gander:2008}, like the time-dependent Schr\"odinger equation.  We will now show how our variational principle can be used to naturally formulate a parallel-in-time algorithm, and demonstrate its improved domain of convergence with respect to the parareal algorithm for Schr\"odinger evolution of Hydrogen.

\begin{figure}
\centering
\includegraphics[width=8cm]{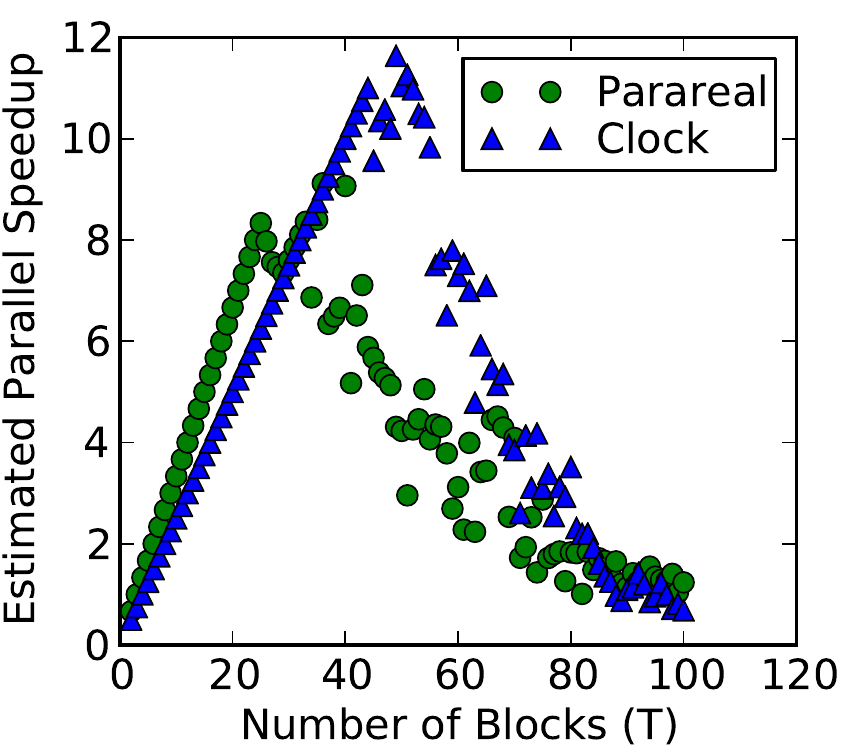}
\caption{In simulating the nuclear dynamics of the Hydrogen molecule, the clock formulation not only demonstrates higher peak parallel speedup compared to parareal, but more robust speedup for longer total evolution times.  This is an isoaccuracy comparison in that all points are converged to an identical level of accuracy corresponding to a fine timestep of $dt=0.015$. The results have been averaged for 10 consecutive evolutions of the specified number of blocks $T$.  For small times the parareal algorithm has a slight advantage due to reduced overhead, but as the total evolution time increases it is less robust to the diminishing quality of the preconditioner.  The non-monotonic nature of the speedup results from the preconditioner having variable quality depending on the dynamics of the system.}
\label{fig:paritime}
\end{figure}

Starting from the TEDVP, minimization under the constraint that the initial state is fixed yields a Hermitian positive-definite, block-tridiagonal linear equation of the form
\begin{equation}
\mathcal{R}^f\ket{\Phi} = \ket{\Lambda}
\end{equation}
where $\ket{\Phi}$(to be solved for) contains the full evolution of the system and $\ket{\Lambda}$ specifies the initial condition such that

\begin{equation}
\ket{\Lambda} = \left(
\begin{array}{cccccc}
\frac{1}{2}\ket{\Psi_0} & 0 & 0 & 0 & ... & 0
\end{array} \right)^T
\end{equation}

The linear clock operator, $\mathcal{R}^i$, is similar to before, where now we distinguish between a clock built from a coarse operator, $\mathcal{R}^c$, from a clock built from a fine operator, $\mathcal{R}^f$, as

\begin{equation}
\mathcal{R}^i=\left(
\begin{array}{cccccccc}
  I & -\frac{1}{2} U^{i\dagger}_{0}& 0 &...\\
  -\frac{1}{2}U^i_{0} & I & -\frac{1}{2}U^{i\dagger}_{1} &\\
  0       & -\frac{1}{2}U^i_{1} & I\\
   &\vdots &\ddots &\ddots&&\\ 
  &&- \frac{1}{2}U^i_{T-2} & I & -\frac{1}{2}U^{i\dagger}_{T-1}  \\
  &&0& -\frac{1}{2}U^i_{T-1} & \frac{1}{2} I\\
\end{array}\right)
\end{equation}

The spectrum of this operator is positive-definite and admits $T$ distinct eigenvalues, each of which is $N$-fold degenerate. The conjugate gradient algorithm can be used to solve for $\ket{\Phi}$, converging at-worst in a number of steps which is equal to the number of distinct eigenvalues\cite{Hestenes:1952,Faber:1984}. Thus application of the conjugate gradient algorithm to this problem is able to converge requiring at most $T$ applications of the linear clock operator $\mathcal{R}^f$. This approach on its own yields no parallel speedup.  However, the use of a well-chosen preconditioner can greatly accelerate the convergence of the conjugate gradient algorithm\cite{Eisenstat:1981}.

If one uses an approximate propagation performed in serial, $\mathcal{R}^c$, which is much cheaper to perform than the exact propagation, as a preconditioning step to the conjugate gradient solve, the algorithm can converge in far fewer steps than $T$ and a parallel-in-time speedup can be achieved.   The problem being solved in this case for the clock construction is given by
\begin{equation}
(\mathcal{R}^c)^{-1}\mathcal{R}^f\ket{\Phi} = (\mathcal{R}^c)^{-1} \ket{\Lambda}
\end{equation}

To clarify and compare with existing methods, we now introduce an example from chemistry.  The nuclear quantum dynamics of the Hydrogen molecule in its ground electronic state can be modeled by the Hamiltonian
\begin{equation}
H=-\frac{\hat P^2}{2m} + D \left(e^{-2 \beta \hat X} - 2 e^{- \beta \hat X} \right)
\end{equation}
where $m=918.5$, $\beta=0.9374$,  and $D=0.164$ atomic units\cite{Makri:1993}.  The initial state of our system is a gaussian wavepacket with a width corresponding to the ground state of the harmonic approximation to this potential, displaced $-0.1$ \AA\  from the equilibrium position.  To avoid the storage associated with the propagator of this system and mimic the performance of our algorithm on a larger system, we use the symmetric matrix-free split-operator Fourier transform method(SOFT) to construct block propagators\cite{Feit:1982}. This method is unconditionally stable, accurate to third order in the time step $dt$, and may be written as
\begin{equation}
U_{SOFT}(t+dt, t) = e^{-i V(\hat X) dt/2} F^{-1}e^{-i \hat P^2/(2m) dt}F e^{-i V(\hat X) dt/2}
\end{equation}
Here, $F$ and $F^{-1}$ corresponds to the application of the fast Fourier transform (FFT) and its inverse over the wavefunction grid.  The use of the FFT allows each of the exponential operators to be applied trivially to their eigenbasis and as a result the application of the propagator has a cost dominated by the FFT that scales as $O(N \log N)$, where $N$ is the number of grid-points being used to represent $\ket{\psi}$.  For our algorithm, we define a fine propagator, $U^f$, and a coarse propagator, $U^c$ from the SOFT construction, such that for a given number of sub-time steps $k$.
\begin{align}
U^{f} &= U_{SOFT}(t+kdt, t + (k-1)dt)...U_{SOFT}(t+dt, t) \\
U^{c} &= U_{SOFT}(t+kdt, t)
\end{align}

We take for our problem the clock constructed from the fine-propagator and use the solution of the problem built from the coarse propagator as our preconditioner.  In all cases, only the matrix free version of the propagator is used, including in the explicit solution of the coarse propagation.  

From the construction of the coarse and fine propagators, with $T$ processors, up to communication time, the cost of applying the fine clock in parallel and solving the coarse clock in serial require roughly the same amount of computational time.  This is a good approximation in the usual case where the application of the propagators is far more expensive than the communication required. From this, assuming the use of $T$ processors, we define an estimated parallel-in-time speed-up for the computational procedure given by
\begin{equation}
S_{clock} = \frac{T}{2 (N_{f} + N_{c})}
\end{equation}
where $N_{f}$ is the number of applications of the fine-propagator matrix $\mathcal{R}^f$ performed in parallel and $N_{c}$ is the number of serial linear solves using the coarse-propagator matrix $\mathcal{R}^c$ used in preconditioned conjugate gradient.  The factor of 2 originates from the requirement of backwards evolution not present in a standard propagation. The cost of communication overhead as well as the inner-products in the CG algorithm are neglected for this approximate study, assuming they are negligible with respect to the cost of applying the propagator.

The equivalent parallel speedup for the parareal algorithm is given by
\begin{equation}
S_{para} = \frac{T}{N_{f} + N_{c}}
\end{equation}
where $N_{f}$ and $N_{c}$ are now defined for the corresponding parareal operators which are functionally identical to the clock operators without backward time evolution, and thus it lacks the same factor of 2.

As is stated above, in the solution of the linear clock without preconditioning, it is possible to converge the problem in at most $T$ steps, independent of both the choice of physical timestep and the size of the physical system $N$ by using a conjugate gradient method.  However, with the addition of the preconditioner, the choice of timestep and total time simulated can have an effect on the total preconditioned system.  This is because as the coarse (approximate) propagation deviates more severely from the exact solution, the preconditioning of the problem can become quite poor.

This problem exhibits itself in a more extreme way for the parareal algorithm, as the predictor-corrector scheme may start to diverge for very poorly preconditioned system.  This has been seen before in the study of hyperbolic equations\cite{Gander:2008}, and remains true in this case for the evolution of the Schr\"odinger equation.  The construction derived from the clock is more robust and is able to achieve parallel-in-time speedup for significantly longer times. This marks an improvement over the current standard for parallel-in-time evolution of the Schr\"odinger equation. 

To give a concrete example, consider the case where we divide the evolution of the nuclear dynamics of hydrogen into pieces containing $T$ evolution blocks, each of which is constructed from $T$ fine evolutions for a time $dt=0.015$ as is described above.  The dynamics over which we simulate are depicted in Fig. 8.  We average the estimated parallel speedup for 10 time blocks (which we define as the whole time in one construction of the clock) forward and the results are for the speedup are given in Fig. 7.  In this example we see that for small $T$ (or small total evolution times), the reduced overhead of having no backwards evolution yields an advantage for the parareal algorithm.  However as the $T$ increases, the parareal algorithm is less robust to errors in the coarse propagation and performance begins to degrade rapidly.  In these cases, our clock construction demonstrates a clear advantage. It is a topic of current research how to facilitate even longer evolutions in the clock construction.

%
%

\section{Norm Loss as a measure of truncation error}
Conservation of the norm of a wavefunction is often considered a critical property for approximate quantum dynamics, as it is a property of the exact propagator resulting from time-reversal symmetry. However, if norm conservation is enforced in a propagation which does not span the complete Hilbert-space, one must account for the components of the wavefunction that would have evolved into the space not included in the computation. It's not immediately clear how population density which attempts to leave the selected subspace should be folded back into the system without being able to simulate the exact dynamics. This problem is sometimes glossed over with the use of exponential propagators that are guaranteed to produce norm-conserving dynamics on a projected Hamiltonian. Some more sophisticated approaches adjust the configuration space in an attempt to mitigate the error\cite{Westermann:2012dq}.

\begin{figure}
\centering
\includegraphics[width=8cm]{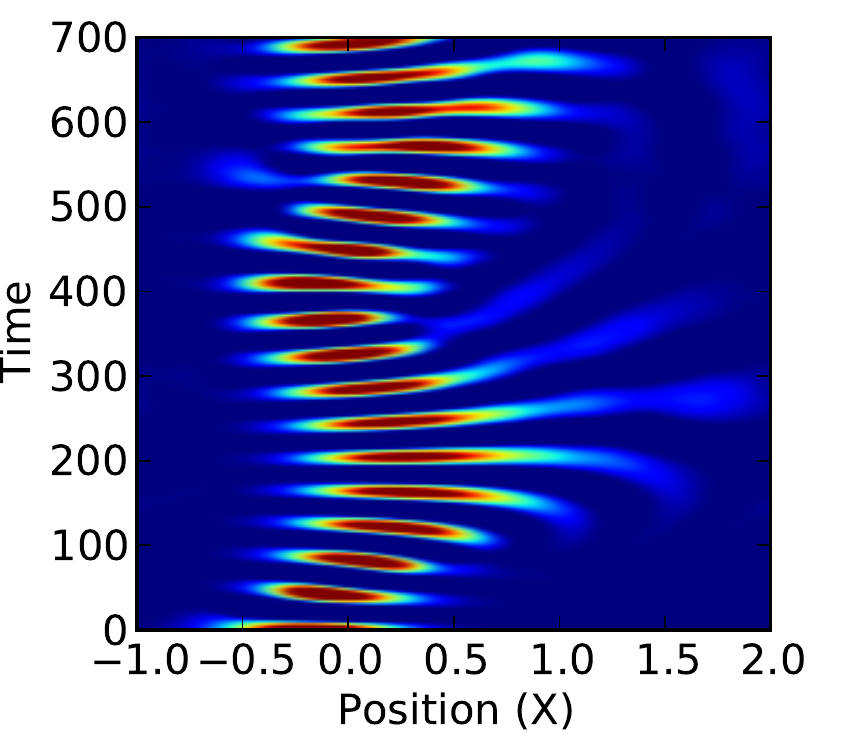}
\caption{The real part of the Clock solution for the nuclear dynamics of the hydrogen molecule oscillates around an equilibrium bond length as expected, eliciting diverse interference patterns due to anharmonicity.  Both time and position are given in atomic units and the color indicates the value of the real part of the waveform at that space-time point.}
\label{fig:clockwave}
\end{figure}

This discrepancy is at the center of the difference between the approximate dynamics derived from the discrete
variational principle here and the approximate dynamics derived from the McLachlan variational
principle such as the multi-configurational time-dependent Hartree method.  Mathematically, this results
from the non-commutativity of the exponential map and projection operator defined above. That is
$P_{\mathcal{B}_i(t)} \mathcal{T} \left( e^{-i \int dt' H(t')} \right) P_{\mathcal{B}_i(t)} \neq \mathcal{T} \left( \exp \left(-i \int dt' P_{\mathcal{B}_i(t') } H(t') P_{\mathcal{B}_i(t')} \right) \right)$ for a Hermitian operator $H$.  In an approximate method derived from the McLachlan or any of the other differential time-dependent variational principles, the projection is performed on the Hamiltonian.  As the projection of any Hermitian operator yields another Hermitian operator, the dynamics generated from the projection are guaranteed to be unitary if a sufficiently accurate exponential propagator is used.  Conversely, projection of a unitary operator, as prescribed by the TEDVP, does not always yield a unitary operator.  Thus for an approximate basis, one expects norm conservation to be violated, where the degree of violation is related to the severity of the configuration space truncation error. This leads us to define a metric of truncation error which we term the instantaneous norm loss.  We define this as
\begin{equation}
  \mathcal{N}_1(t) = 1 - \| P_{\mathcal{B}_i} U_t P_{\mathcal{B}_i} \ket{\tilde \Psi_t} \|^2
\end{equation}
where $\ket{\tilde \Psi_t}$ is always assumed to be normalized, which in practice means that a renormalization is used after each time step here.  However, as we proved above, in the limit of a short time step, with
dynamics generated by a Hamiltonian, the TEDVP must contain essentially the same content as the McLachlan
variational principle.  For this reason, we propose an additional metric which is given by
\begin{equation}
  \mathcal{N}_2(t) = \| (H(t) - P_{\mathcal{B}_i} H(t) P_{\mathcal{B}_i} )\ket{\tilde \Psi_t} \|^2 dt^2
  = \|V(t) \ket{\tilde \Psi_t} \|^2 dt^2
\end{equation}
Where $H(t)$ is the physical Hamiltonian.  This is motivated by appealing to the McLachlan variational
principle and substituting from the exact Schr\"odinger equation that $i \partial_t = H \ket{\Psi_t}$
where $H$ is the full (non-projected) Hamiltonian. By defining $V(t) = (H(t) - P_{\mathcal{B}_i} H(t) P_{\mathcal{B}_i})$ , we recognize this as a pertubation theory estimate
of the error resulting from the configuration basis truncation at a given point in time.

To examine the quality of these metrics and to better understand the consequences of the non-commutativity of
the exponential map and projection, we return to the sample spin system.
In this case, we choose a basis for the propagation space based entirely on the initial state, and do not allow it to change dynamically in time as before.  We perform simulations in the space of single excitations (S) from the initial state, double excitations (SD), and in the full Hilbert space (Ex).
Dynamics from the TEDVP are generated by first building the exact propagator then projecting to the desired basis set while dynamics from the McLachlan variational principle (MVP) are modeled by projecting the Hamiltonian into the target basis set and exponentiating.  A renormalization is used after each time step in the first case.  Although one could perform the simulation with a timestep where timestep error is negligible, we remove this component of the calculation for this example by making the Hamiltonian time-independent.  This allows direct analysis of the effect of timestep on non-commutativity deviations.

In Fig. 6 we show the dynamics of the Vanadium spin complex for the two approximate truncation levels (S and SD) with both methods and their associated error metrics($\mathcal{N}_1(t)$ and $\mathcal{N}_2(t)$).
Deviations in the qualitative features of the observable occur after even the first peak of
the proposed metrics.  In this particular simulation, the configuration interaction with
singles and doubles spans all but one state in the full Hilbert space.  The lack of this one
state results in the large qualitative errors present, associated with the first and subsequent peaks
present in these error metrics.  The impact of later peaks is more difficult to discern, due to error in the wavefunctions, which accumulates as the propagation proceeds.

As predicted by the connection between the TEDVP and the McLachlan variational principle,
while $\mathcal{N}_1$ and $\mathcal{N}_2$ are not identical for each case, in the short time limit they 
yield extremely similar information, which is highlighted in Fig. 6 displaying
the resulting longer time dynamics for a time step of $dt=0.01$.  In Fig. 9, however we explore the effects of a significantly larger time step and begin to discern the result of the non-commutativity discussed
here.  Recalling that because the Hamiltonian is time-independent in this case, the propagator used
is numerically exact in both instances, so this effect is not a result of what would be traditionally
called time step error, resulting from intrinsic errors of an integrator.  Interestingly, it is observed that $\mathcal{N}_1(t)$ begins to decay to a nearly
constant value.  This occurs because the action of projection after exponentiation breaks
the degeneracy of the spectrum of the unitary operator, resulting in eigenvalues with norms different than $1$.
As a result, repeated action and renormalization of the operator is analogous to a
power method for finding the eigenvector associated with the largest eigenvalue.  This
effect is exacerbated by taking long time steps. 

\begin{figure}[ht]
\centering
\includegraphics[width=8cm]{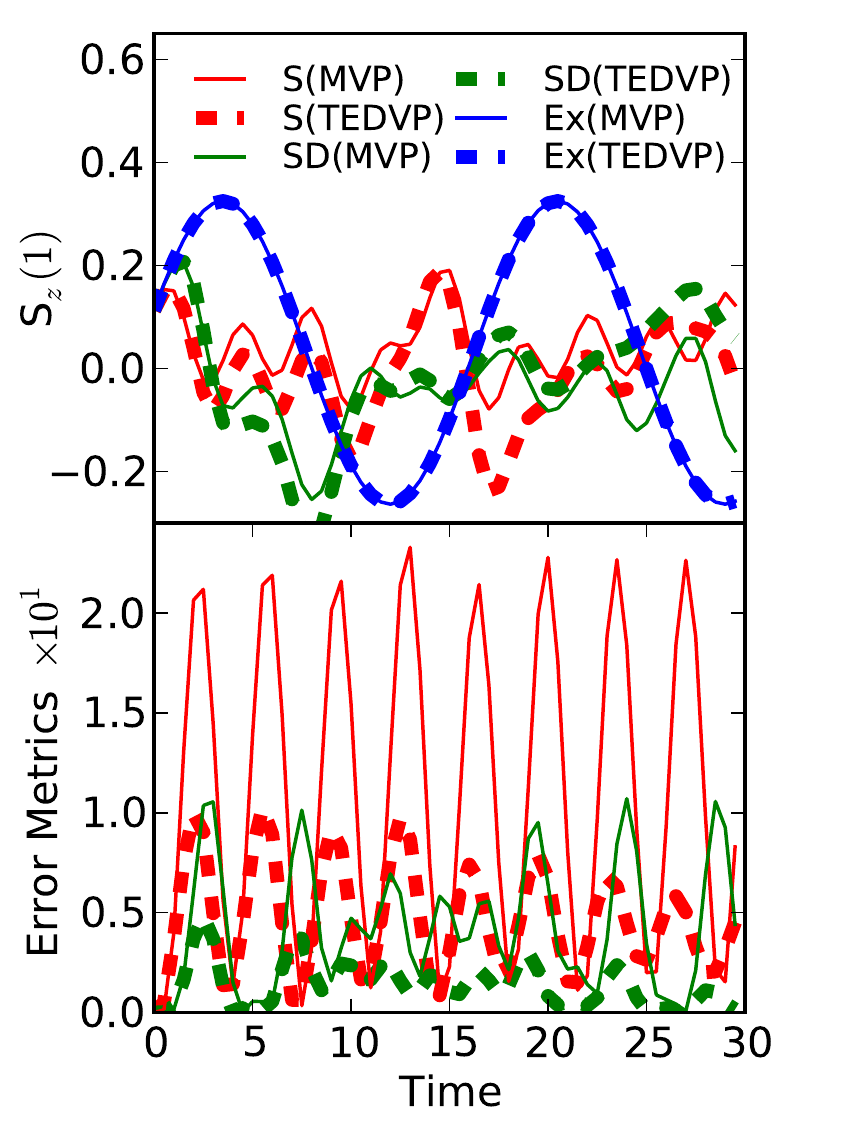}
\caption{The dynamics of a spin observable in the Vanadium complex at a large time step, $dt=0.5 K^{-1}$, shows significant differences resulting from the non-commutativity of projection and exponentiation in dynamics generated by the time embedded discrete variational principle(TEDVP) (dashed lines) and the McLachlan variational principle(MVP) (solid lines).  Results are shown for propagations restricted to the space of single excitations from the initial state(S), double excitations (SD), and the full space(Ex).  The corresponding error metrics, $\mathcal{N}_1(t)$ for the TEDVP (dashed) and $\mathcal{N}_2(t)$ for the MVP (solid), differ considerably in this case.}
\label{fig:NormLoss5}
\end{figure}

\section{Conclusions}
In this manuscript, we introduce a new variational principle for time-dependent dynamics 
inspired by the Feynman clock originally employed for quantum computation. Unlike other 
previously-proposed variational principles, the proposed TEDVP approach involves the 
solution of an eigenvalue problem for the entire time propagation. This perspective allows for 
readily employing many of the powerful truncation techniques from quantum chemistry and 
condensed-matter physics, that have been developed for the exact diagonalization problem. 
We show how this formulation naturally leads to a parallel-in-time algorithm and demonstrate its improved robustness with respect to existing methods.  We introduce two novel error metrics for the TEDVP that allow one to 
characterize the basis approximations involved. The features of the method were demonstrated by 
simulating the dynamics of a Hydrogen molecule and a molecular effective spin Hamiltonian. Further research  directions include the use of other approximate techniques for the time dynamics such as the 
use of perturbation theory \cite{Helgaker:2002} or coupled-cluster 
approaches\cite{Kvaal:2012cr}, and further enhancement of parallel-in-time dynamics.

\subsection{Acknowledgments}
We would like to acknowledge David Tempel for his
valuable comments on the manuscript. J.M. is supported by the DOE Computational
Science Graduate Fellowship under grant number DE-FG02-97ER25308.
A.A.-G. and J.P. thank the National Science Foundation for their support under award number CHE-1152291.  
A.A.-G. receives support from the HRL Laboratories, LLC under award number
M1144-201167-DS and UCSD under grant number FA9550-12-1-0046. Research sponsored by United States Department of Defense. The views and conclusions contained in this document are those of the authors and should not be interpreted as representing the official policies, either expressly or implied, of the U.S. Government. We also thank the Corning Foundation, the Camille and Henry Dreyfus Foundation, and the Alfred P. Sloan Foundation for their continuing support of this research. 

\subsection{Appendix: On the general construction of eigenvalue problems from dynamics}
We have presented one path for constructing eigenvalue problems from quantum
dynamics problems so far, however it is instructive to illuminate precisely
which part of our procedure allowed this.  To do this, we will slightly
generalize the Floquet-type Hamiltonians and demonstrate that the time embedding
was the crucial feature that allows one to recast a dynamics problem as
an eigenvalue problem.  This is in addition to the choice to work in an
integrated framework, which we will show simply allows for a convenient choice of basis.

Recall the definition of a Floquet-type Hamiltonian given by
\begin{equation}
 F(t) = H(t) - i \partial_t
\end{equation}
If one considers a finite time evolution for a length of time $T$, 
this operator is Hermitian in the basis of Fourier functions states given by
\begin{equation}
 \ket{\Psi_{nj}(t)} = \ket{\Phi_j} \ket{n} = \ket{\Phi_j} e^{2 \pi i n t/T}
\end{equation}
where $\ket{\Phi_j}$ is a time-independent state of the physical system when considering
the generalized inner-product 
\begin{equation}
 \bra{n'} \braket{\Phi_i}{\Phi_j} \ket{n} 
    = \frac{\braket{\Phi_i}{\Phi_j}}{T} \int_0^T dt' e^{- 2\pi i n' t'/T} e^{2 \pi i n t'/T} 
\end{equation}
This is the the generalized Hilbert space first introduced by Sambe~\cite{Sambe:1973} and generalized by Howland~\cite{Howland:1974}. Because this operator is Hermitian in this basis, by noting the similarity to the Lagrange Variational principle
\begin{equation}
 \delta L = \delta \int_0^T dt \bra{\Psi(t)} F \ket{\Psi(t)} = 0 
\end{equation}
minimization of $L$ on this linear basis of Fourier states yields a Hermitian eigenvalue problem.  Thus the time evolution can be reconstructed by solving the full time-independent eigenvalue 
problem in this basis, or by constructing a surrogate evolution operator as in the $(t,t')$
formalism of Peskin and Moiseyev~\cite{Peskin:1993}.  The use of Fourier basis states to express time dependence is natural given the form of the operator $F$.  That is, matrix elements of
the derivative operator $\partial_t$ have a trivial analytic expression
in this basis.   This represents  the same general idea we have been discussing here, which is to consider states in a system-time Hilbert space. However, as the solution of this variational problem will may yield a stationary point rather than a true minimum~\cite{McLachlan:1964}, ground state techniques are not appropriate for this particular operator.  Moreover, this operator is not in general Hermitian when considering arbitrary basis functions of time. 
For example arbitrary choices of plane waves not corresponding to the traditional Fourier basis will yield a non-Hermitian operator.  The operator $F' = (1/2)(F + F^\dagger)$, which has been used in the past for the construction of approximate dynamics~\cite{Shalashilin:2008}, is Hermitian, however it still suffers from a pathology that the optimal solution represents a stationary point rather than a minimum. However, the operator
\begin{equation}
 G = F^\dagger F = (H(t) - i \partial_t)^\dagger(H(t) - i \partial_t)
\end{equation}
is always Hermitian and positive semi-definite by construction.  
Thus one can expand the system-time Hilbert space in any linear basis
of time, and the optimal path in that space will be the ground state
eigenvector of the operator $G$ utilizing the above generalized inner-product, assuming we have broken degeneracy by introducing the correct initial state.  This can be viewed as an application of the McLachlan time dependent variational principle.  From this, we see it is the expression systems in a system-time Hilbert space which allows
for the eigenvalue problem construction.  Moreover, we note that this is not limited
to the use of Fourier time basis states, and that many more expressions
of time dependence may be utilized to construct an eigenvalue problem within this framework.

One may ask then, what was the objective of working in the an integrated formalism, defined using unitary operators rather than differential operators.  To see this, consider evaluating  the system at a point in time with the Fourier construction above.
One has to expand the system in all time basis functions and evaluate them at a time $t$.  
When one usually considers time, however, they are thinking of a parametric
construction where a number $t$ simply labels a specific state of a system.  Embedding
into the system-time Hilbert space with this idea would be most naturally
expressed as delta-functions.  However matrix elements of $\partial_t$ on this basis
can be difficult to construct.  In the ancillary time system framework used in the TEDVP, however, time is easily expressed as a discrete parametric variable.  One might also
consider the use of discretized derivatives in the operator $G$.  However, balancing
the errors in numerical derivatives and the increasing difficultly of solving the problem
with the number of simultaneously stored time steps can be practically difficult.

\bibliographystyle{apsrev4-1}
\bibliography{ClockArxiv}

\begin{thebibliography}{53}%
\makeatletter
\providecommand \@ifxundefined [1]{%
 \@ifx{#1\undefined}
}%
\providecommand \@ifnum [1]{%
 \ifnum #1\expandafter \@firstoftwo
 \else \expandafter \@secondoftwo
 \fi
}%
\providecommand \@ifx [1]{%
 \ifx #1\expandafter \@firstoftwo
 \else \expandafter \@secondoftwo
 \fi
}%
\providecommand \natexlab [1]{#1}%
\providecommand \enquote  [1]{``#1''}%
\providecommand \bibnamefont  [1]{#1}%
\providecommand \bibfnamefont [1]{#1}%
\providecommand \citenamefont [1]{#1}%
\providecommand \href@noop [0]{\@secondoftwo}%
\providecommand \href [0]{\begingroup \@sanitize@url \@href}%
\providecommand \@href[1]{\@@startlink{#1}\@@href}%
\providecommand \@@href[1]{\endgroup#1\@@endlink}%
\providecommand \@sanitize@url [0]{\catcode `\\12\catcode `\$12\catcode
  `\&12\catcode `\#12\catcode `\^12\catcode `\_12\catcode `\%12\relax}%
\providecommand \@@startlink[1]{}%
\providecommand \@@endlink[0]{}%
\providecommand \url  [0]{\begingroup\@sanitize@url \@url }%
\providecommand \@url [1]{\endgroup\@href {#1}{\urlprefix }}%
\providecommand \urlprefix  [0]{URL }%
\providecommand \Eprint [0]{\href }%
\providecommand \doibase [0]{http://dx.doi.org/}%
\providecommand \selectlanguage [0]{\@gobble}%
\providecommand \bibinfo  [0]{\@secondoftwo}%
\providecommand \bibfield  [0]{\@secondoftwo}%
\providecommand \translation [1]{[#1]}%
\providecommand \BibitemOpen [0]{}%
\providecommand \bibitemStop [0]{}%
\providecommand \bibitemNoStop [0]{.\EOS\space}%
\providecommand \EOS [0]{\spacefactor3000\relax}%
\providecommand \BibitemShut  [1]{\csname bibitem#1\endcsname}%
\let\auto@bib@innerbib\@empty
\bibitem [{\citenamefont {Feynman}(1982)}]{Feynman:1982}%
  \BibitemOpen
  \bibfield  {author} {\bibinfo {author} {\bibfnamefont {R.}~\bibnamefont
  {Feynman}},\ }\href {http://dx.doi.org/10.1007/BF02650179} {\bibfield
  {journal} {\bibinfo  {journal} {Int. J. Theor. Phys.}\ }\textbf {\bibinfo
  {volume} {21}},\ \bibinfo {pages} {467} (\bibinfo {year} {1982})}\BibitemShut
  {NoStop}%
\bibitem [{\citenamefont {Feynman}(1985)}]{Feynman:1985}%
  \BibitemOpen
  \bibfield  {author} {\bibinfo {author} {\bibfnamefont {R.~P.}\ \bibnamefont
  {Feynman}},\ }\href {\doibase 10.1364/ON.11.2.000011} {\bibfield  {journal}
  {\bibinfo  {journal} {Opt. News}\ }\textbf {\bibinfo {volume} {11}},\
  \bibinfo {pages} {11} (\bibinfo {year} {1985})}\BibitemShut {NoStop}%
\bibitem [{\citenamefont {Lippmann}\ and\ \citenamefont
  {Schwinger}(1950)}]{Lippmann:1950ly}%
  \BibitemOpen
  \bibfield  {author} {\bibinfo {author} {\bibfnamefont {B.~A.}\ \bibnamefont
  {Lippmann}}\ and\ \bibinfo {author} {\bibfnamefont {J.}~\bibnamefont
  {Schwinger}},\ }\href {\doibase 10.1103/PhysRev.79.469} {\bibfield  {journal}
  {\bibinfo  {journal} {Phys. Rev.}\ }\textbf {\bibinfo {volume} {79}},\
  \bibinfo {pages} {469} (\bibinfo {year} {1950})}\BibitemShut {NoStop}%
\bibitem [{\citenamefont {Heller}(1976)}]{Heller:1976ve}%
  \BibitemOpen
  \bibfield  {author} {\bibinfo {author} {\bibfnamefont {E.~J.}\ \bibnamefont
  {Heller}},\ }\href {\doibase 10.1063/1.431911} {\bibfield  {journal}
  {\bibinfo  {journal} {J. Chem. Phys.}\ }\textbf {\bibinfo {volume} {64}},\
  \bibinfo {pages} {63} (\bibinfo {year} {1976})}\BibitemShut {NoStop}%
\bibitem [{\citenamefont {Kerman}\ and\ \citenamefont
  {Koonin}(1976)}]{Kerman:1976bh}%
  \BibitemOpen
  \bibfield  {author} {\bibinfo {author} {\bibfnamefont {A.}~\bibnamefont
  {Kerman}}\ and\ \bibinfo {author} {\bibfnamefont {S.}~\bibnamefont
  {Koonin}},\ }\href {\doibase 10.1016/0003-4916(76)90065-8} {\bibfield
  {journal} {\bibinfo  {journal} {Annals of Physics}\ }\textbf {\bibinfo
  {volume} {100}},\ \bibinfo {pages} {332 } (\bibinfo {year}
  {1976})}\BibitemShut {NoStop}%
\bibitem [{\citenamefont {Jackiw}\ and\ \citenamefont
  {Kerman}(1979)}]{Jackiw:1979zr}%
  \BibitemOpen
  \bibfield  {author} {\bibinfo {author} {\bibfnamefont {R.}~\bibnamefont
  {Jackiw}}\ and\ \bibinfo {author} {\bibfnamefont {A.}~\bibnamefont
  {Kerman}},\ }\href {\doibase 10.1016/0375-9601(79)90151-8} {\bibfield
  {journal} {\bibinfo  {journal} {Phys. Lett. A}\ }\textbf {\bibinfo {volume}
  {71}},\ \bibinfo {pages} {158 } (\bibinfo {year} {1979})}\BibitemShut
  {NoStop}%
\bibitem [{\citenamefont {Balian}\ and\ \citenamefont
  {V\'en\'eroni}(1981)}]{Balian:1981ys}%
  \BibitemOpen
  \bibfield  {author} {\bibinfo {author} {\bibfnamefont {R.}~\bibnamefont
  {Balian}}\ and\ \bibinfo {author} {\bibfnamefont {M.}~\bibnamefont
  {V\'en\'eroni}},\ }\href {\doibase 10.1103/PhysRevLett.47.1353} {\bibfield
  {journal} {\bibinfo  {journal} {Phys. Rev. Lett.}\ }\textbf {\bibinfo
  {volume} {47}},\ \bibinfo {pages} {1353} (\bibinfo {year}
  {1981})}\BibitemShut {NoStop}%
\bibitem [{\citenamefont {Deumens}\ \emph {et~al.}(1994)\citenamefont
  {Deumens}, \citenamefont {Diz}, \citenamefont {Longo},\ and\ \citenamefont
  {\"Ohrn}}]{Deumens:1994qf}%
  \BibitemOpen
  \bibfield  {author} {\bibinfo {author} {\bibfnamefont {E.}~\bibnamefont
  {Deumens}}, \bibinfo {author} {\bibfnamefont {A.}~\bibnamefont {Diz}},
  \bibinfo {author} {\bibfnamefont {R.}~\bibnamefont {Longo}}, \ and\ \bibinfo
  {author} {\bibfnamefont {Y.}~\bibnamefont {\"Ohrn}},\ }\href {\doibase
  10.1103/RevModPhys.66.917} {\bibfield  {journal} {\bibinfo  {journal} {Rev.
  Mod. Phys.}\ }\textbf {\bibinfo {volume} {66}},\ \bibinfo {pages} {917}
  (\bibinfo {year} {1994})}\BibitemShut {NoStop}%
\bibitem [{\citenamefont {Poulsen}(2011)}]{Poulsen:2011kl}%
  \BibitemOpen
  \bibfield  {author} {\bibinfo {author} {\bibfnamefont {J.~A.}\ \bibnamefont
  {Poulsen}},\ }\href {\doibase 10.1063/1.3519637} {\bibfield  {journal}
  {\bibinfo  {journal} {J. Chem. Phys.}\ }\textbf {\bibinfo {volume} {134}},\
  \bibinfo {eid} {034118} (\bibinfo {year} {2011})}\BibitemShut {NoStop}%
\bibitem [{\citenamefont {Haegeman}\ \emph {et~al.}(2011)\citenamefont
  {Haegeman}, \citenamefont {Cirac}, \citenamefont {Osborne}, \citenamefont
  {Pi\ifmmode~\check{z}\else \v{z}\fi{}orn}, \citenamefont {Verschelde},\ and\
  \citenamefont {Verstraete}}]{Haegeman:2011tg}%
  \BibitemOpen
  \bibfield  {author} {\bibinfo {author} {\bibfnamefont {J.}~\bibnamefont
  {Haegeman}}, \bibinfo {author} {\bibfnamefont {J.~I.}\ \bibnamefont {Cirac}},
  \bibinfo {author} {\bibfnamefont {T.~J.}\ \bibnamefont {Osborne}}, \bibinfo
  {author} {\bibfnamefont {I.}~\bibnamefont {Pi\ifmmode~\check{z}\else
  \v{z}\fi{}orn}}, \bibinfo {author} {\bibfnamefont {H.}~\bibnamefont
  {Verschelde}}, \ and\ \bibinfo {author} {\bibfnamefont {F.}~\bibnamefont
  {Verstraete}},\ }\href {\doibase 10.1103/PhysRevLett.107.070601} {\bibfield
  {journal} {\bibinfo  {journal} {Phys. Rev. Lett.}\ }\textbf {\bibinfo
  {volume} {107}},\ \bibinfo {pages} {070601} (\bibinfo {year}
  {2011})}\BibitemShut {NoStop}%
\bibitem [{\citenamefont {Milfeld}\ and\ \citenamefont
  {Wyatt}(1983)}]{Milfeld:1983}%
  \BibitemOpen
  \bibfield  {author} {\bibinfo {author} {\bibfnamefont {K.~F.}\ \bibnamefont
  {Milfeld}}\ and\ \bibinfo {author} {\bibfnamefont {R.~E.}\ \bibnamefont
  {Wyatt}},\ }\href {\doibase 10.1103/PhysRevA.27.72} {\bibfield  {journal}
  {\bibinfo  {journal} {Phys. Rev. A}\ }\textbf {\bibinfo {volume} {27}},\
  \bibinfo {pages} {72} (\bibinfo {year} {1983})}\BibitemShut {NoStop}%
\bibitem [{\citenamefont {Autler}\ and\ \citenamefont
  {Townes}(1955)}]{Autler:1955}%
  \BibitemOpen
  \bibfield  {author} {\bibinfo {author} {\bibfnamefont {S.~H.}\ \bibnamefont
  {Autler}}\ and\ \bibinfo {author} {\bibfnamefont {C.~H.}\ \bibnamefont
  {Townes}},\ }\href {\doibase 10.1103/PhysRev.100.703} {\bibfield  {journal}
  {\bibinfo  {journal} {Phys. Rev.}\ }\textbf {\bibinfo {volume} {100}},\
  \bibinfo {pages} {703} (\bibinfo {year} {1955})}\BibitemShut {NoStop}%
\bibitem [{\citenamefont {Dion}\ and\ \citenamefont
  {Hirschfelder}(2007)}]{Dion:2007}%
  \BibitemOpen
  \bibfield  {author} {\bibinfo {author} {\bibfnamefont {D.~R.}\ \bibnamefont
  {Dion}}\ and\ \bibinfo {author} {\bibfnamefont {J.~O.}\ \bibnamefont
  {Hirschfelder}},\ }\enquote {\bibinfo {title} {Time-dependent perturbation of
  a two-state quantum system by a sinusoidal field},}\ in\ \href {\doibase
  10.1002/9780470142547.ch5} {\emph {\bibinfo {booktitle} {Adv. Chem. Phys.}}}\
  (\bibinfo  {publisher} {John Wiley and Sons, Inc.},\ \bibinfo {year} {2007})\
  pp.\ \bibinfo {pages} {265--350}\BibitemShut {NoStop}%
\bibitem [{\citenamefont {Peskin}\ and\ \citenamefont
  {Moiseyev}(1993)}]{Peskin:1993}%
  \BibitemOpen
  \bibfield  {author} {\bibinfo {author} {\bibfnamefont {U.}~\bibnamefont
  {Peskin}}\ and\ \bibinfo {author} {\bibfnamefont {N.}~\bibnamefont
  {Moiseyev}},\ }\href {\doibase 10.1063/1.466058} {\bibfield  {journal}
  {\bibinfo  {journal} {J. Chem. Phys.}\ }\textbf {\bibinfo {volume} {99}},\
  \bibinfo {pages} {4590} (\bibinfo {year} {1993})}\BibitemShut {NoStop}%
\bibitem [{\citenamefont {Nielsen}\ and\ \citenamefont
  {Chuang}(2000)}]{Nielsen:2000}%
  \BibitemOpen
  \bibfield  {author} {\bibinfo {author} {\bibfnamefont {M.}~\bibnamefont
  {Nielsen}}\ and\ \bibinfo {author} {\bibfnamefont {I.}~\bibnamefont
  {Chuang}},\ }\href {http://books.google.com/books?id=65FqEKQOfP8C} {\emph
  {\bibinfo {title} {Quantum Computation and Quantum Information}}},\ Cambridge
  Series on Information and the Natural Sciences\ (\bibinfo  {publisher}
  {Cambridge University Press},\ \bibinfo {year} {2000})\BibitemShut {NoStop}%
\bibitem [{\citenamefont {Aspuru-Guzik}\ \emph {et~al.}(2005)\citenamefont
  {Aspuru-Guzik}, \citenamefont {Dutoi}, \citenamefont {Love},\ and\
  \citenamefont {Head-Gordon}}]{Aspuru:2005}%
  \BibitemOpen
  \bibfield  {author} {\bibinfo {author} {\bibfnamefont {A.}~\bibnamefont
  {Aspuru-Guzik}}, \bibinfo {author} {\bibfnamefont {A.~D.}\ \bibnamefont
  {Dutoi}}, \bibinfo {author} {\bibfnamefont {P.~J.}\ \bibnamefont {Love}}, \
  and\ \bibinfo {author} {\bibfnamefont {M.}~\bibnamefont {Head-Gordon}},\
  }\href {\doibase 10.1126/science.1113479} {\bibfield  {journal} {\bibinfo
  {journal} {Science}\ }\textbf {\bibinfo {volume} {309}},\ \bibinfo {pages}
  {1704} (\bibinfo {year} {2005})}\BibitemShut {NoStop}%
\bibitem [{\citenamefont {Kassal}\ \emph {et~al.}(2008)\citenamefont {Kassal},
  \citenamefont {Jordan}, \citenamefont {Love}, \citenamefont {Mohseni},\ and\
  \citenamefont {Aspuru-Guzik}}]{Kassal:2008}%
  \BibitemOpen
  \bibfield  {author} {\bibinfo {author} {\bibfnamefont {I.}~\bibnamefont
  {Kassal}}, \bibinfo {author} {\bibfnamefont {S.~P.}\ \bibnamefont {Jordan}},
  \bibinfo {author} {\bibfnamefont {P.~J.}\ \bibnamefont {Love}}, \bibinfo
  {author} {\bibfnamefont {M.}~\bibnamefont {Mohseni}}, \ and\ \bibinfo
  {author} {\bibfnamefont {A.}~\bibnamefont {Aspuru-Guzik}},\ }\href {\doibase
  10.1073/pnas.0808245105} {\bibfield  {journal} {\bibinfo  {journal} {Proc.
  Natl. Acad. Sci. USA}\ }\textbf {\bibinfo {volume} {105}},\ \bibinfo {pages}
  {18681} (\bibinfo {year} {2008})}\BibitemShut {NoStop}%
\bibitem [{\citenamefont {Wang}\ \emph {et~al.}(2008)\citenamefont {Wang},
  \citenamefont {Kais}, \citenamefont {Aspuru-Guzik},\ and\ \citenamefont
  {Hoffmann}}]{Wang:2008}%
  \BibitemOpen
  \bibfield  {author} {\bibinfo {author} {\bibfnamefont {H.}~\bibnamefont
  {Wang}}, \bibinfo {author} {\bibfnamefont {S.}~\bibnamefont {Kais}}, \bibinfo
  {author} {\bibfnamefont {A.}~\bibnamefont {Aspuru-Guzik}}, \ and\ \bibinfo
  {author} {\bibfnamefont {M.~R.}\ \bibnamefont {Hoffmann}},\ }\href {\doibase
  10.1039/B804804E} {\bibfield  {journal} {\bibinfo  {journal} {Phys. Chem.
  Chem. Phys.}\ }\textbf {\bibinfo {volume} {10}},\ \bibinfo {pages} {5388}
  (\bibinfo {year} {2008})}\BibitemShut {NoStop}%
\bibitem [{\citenamefont {Kassal}\ and\ \citenamefont
  {Aspuru-Guzik}(2009)}]{Kassal:2009}%
  \BibitemOpen
  \bibfield  {author} {\bibinfo {author} {\bibfnamefont {I.}~\bibnamefont
  {Kassal}}\ and\ \bibinfo {author} {\bibfnamefont {A.}~\bibnamefont
  {Aspuru-Guzik}},\ }\href {\doibase 10.1063/1.3266959} {\bibfield  {journal}
  {\bibinfo  {journal} {J. Chem. Phys.}\ }\textbf {\bibinfo {volume} {131}},\
  \bibinfo {eid} {224102} (\bibinfo {year} {2009})}\BibitemShut {NoStop}%
\bibitem [{\citenamefont {Dirac}(1930)}]{Dirac:1930}%
  \BibitemOpen
  \bibfield  {author} {\bibinfo {author} {\bibfnamefont {P.~A.~M.}\
  \bibnamefont {Dirac}},\ }\href {\doibase 10.1017/S0305004100016108}
  {\bibfield  {journal} {\bibinfo  {journal} {Math. Proc. Cambridge}\ }\textbf
  {\bibinfo {volume} {26}},\ \bibinfo {pages} {376} (\bibinfo {year}
  {1930})}\BibitemShut {NoStop}%
\bibitem [{\citenamefont {Frenkel}(1934)}]{Frenkel:1934}%
  \BibitemOpen
  \bibfield  {author} {\bibinfo {author} {\bibfnamefont {J.}~\bibnamefont
  {Frenkel}},\ }\href@noop {} {\emph {\bibinfo {title} {Wave Mechanics}}}\
  (\bibinfo  {publisher} {Claredon Press, Oxford},\ \bibinfo {year}
  {1934})\BibitemShut {NoStop}%
\bibitem [{\citenamefont {McLachlan}(1964)}]{McLachlan:1964}%
  \BibitemOpen
  \bibfield  {author} {\bibinfo {author} {\bibfnamefont {A.}~\bibnamefont
  {McLachlan}},\ }\href {\doibase 10.1080/00268976400100041} {\bibfield
  {journal} {\bibinfo  {journal} {Mol. Phys.}\ }\textbf {\bibinfo {volume}
  {8}},\ \bibinfo {pages} {39} (\bibinfo {year} {1964})}\BibitemShut {NoStop}%
\bibitem [{\citenamefont {Slater}(1929)}]{Slater:1929}%
  \BibitemOpen
  \bibfield  {author} {\bibinfo {author} {\bibfnamefont {J.~C.}\ \bibnamefont
  {Slater}},\ }\href {\doibase 10.1103/PhysRev.34.1293} {\bibfield  {journal}
  {\bibinfo  {journal} {Phys. Rev.}\ }\textbf {\bibinfo {volume} {34}},\
  \bibinfo {pages} {1293} (\bibinfo {year} {1929})}\BibitemShut {NoStop}%
\bibitem [{\citenamefont {Boys}(1950)}]{Boys:1950}%
  \BibitemOpen
  \bibfield  {author} {\bibinfo {author} {\bibfnamefont {S.~F.}\ \bibnamefont
  {Boys}},\ }\href {\doibase 10.1098/rspa.1950.0036} {\bibfield  {journal}
  {\bibinfo  {journal} {Proc. R. Soc. Lond. A Math. Phys. Sci.}\ }\textbf
  {\bibinfo {volume} {200}},\ \bibinfo {pages} {542} (\bibinfo {year}
  {1950})}\BibitemShut {NoStop}%
\bibitem [{\citenamefont {{Farhi}}\ \emph {et~al.}(2000)\citenamefont
  {{Farhi}}, \citenamefont {{Goldstone}}, \citenamefont {{Gutmann}},\ and\
  \citenamefont {{Sipser}}}]{Farhi:2000}%
  \BibitemOpen
  \bibfield  {author} {\bibinfo {author} {\bibfnamefont {E.}~\bibnamefont
  {{Farhi}}}, \bibinfo {author} {\bibfnamefont {J.}~\bibnamefont
  {{Goldstone}}}, \bibinfo {author} {\bibfnamefont {S.}~\bibnamefont
  {{Gutmann}}}, \ and\ \bibinfo {author} {\bibfnamefont {M.}~\bibnamefont
  {{Sipser}}},\ }\href@noop {} {\  (\bibinfo {year} {2000})},\ \Eprint
  {http://arxiv.org/abs/arXiv:quant-ph/0001106} {arXiv:quant-ph/0001106}
  \BibitemShut {NoStop}%
\bibitem [{\citenamefont {Kitaev}\ \emph {et~al.}(2002)\citenamefont {Kitaev},
  \citenamefont {Shen}, \citenamefont {Vyalyi},\ and\ \citenamefont
  {Vyalyi}}]{Kitaev:2002}%
  \BibitemOpen
  \bibfield  {author} {\bibinfo {author} {\bibfnamefont {A.}~\bibnamefont
  {Kitaev}}, \bibinfo {author} {\bibfnamefont {A.}~\bibnamefont {Shen}},
  \bibinfo {author} {\bibfnamefont {M.}~\bibnamefont {Vyalyi}}, \ and\ \bibinfo
  {author} {\bibfnamefont {N.}~\bibnamefont {Vyalyi}},\ }\href
  {http://books.google.com/books?id=TrMposZZ0MQC} {\emph {\bibinfo {title}
  {Classical and Quantum Computation}}},\ Graduate Studies in Mathematics\
  (\bibinfo  {publisher} {American Mathematical Society},\ \bibinfo {year}
  {2002})\BibitemShut {NoStop}%
\bibitem [{\citenamefont {Coester}\ and\ \citenamefont
  {K\"ummel}(1960)}]{Coester:1960}%
  \BibitemOpen
  \bibfield  {author} {\bibinfo {author} {\bibfnamefont {F.}~\bibnamefont
  {Coester}}\ and\ \bibinfo {author} {\bibfnamefont {H.}~\bibnamefont
  {K\"ummel}},\ }\href {\doibase 10.1016/0029-5582(60)90140-1} {\bibfield
  {journal} {\bibinfo  {journal} {Nucl. Phys.}\ }\textbf {\bibinfo {volume}
  {17}},\ \bibinfo {pages} {477 } (\bibinfo {year} {1960})}\BibitemShut
  {NoStop}%
\bibitem [{\citenamefont {Cizek}(1966)}]{Cizek:1966}%
  \BibitemOpen
  \bibfield  {author} {\bibinfo {author} {\bibfnamefont {J.}~\bibnamefont
  {Cizek}},\ }\href {\doibase 10.1063/1.1727484} {\bibfield  {journal}
  {\bibinfo  {journal} {J. Chem. Phys.}\ }\textbf {\bibinfo {volume} {45}},\
  \bibinfo {pages} {4256} (\bibinfo {year} {1966})}\BibitemShut {NoStop}%
\bibitem [{\citenamefont {Bartlett}\ and\ \citenamefont
  {Musial}(2007)}]{Bartlett:2007}%
  \BibitemOpen
  \bibfield  {author} {\bibinfo {author} {\bibfnamefont {R.~J.}\ \bibnamefont
  {Bartlett}}\ and\ \bibinfo {author} {\bibfnamefont {M.}~\bibnamefont
  {Musial}},\ }\href {\doibase 10.1103/RevModPhys.79.291} {\bibfield  {journal}
  {\bibinfo  {journal} {Rev. Mod. Phys.}\ }\textbf {\bibinfo {volume} {79}},\
  \bibinfo {pages} {291} (\bibinfo {year} {2007})}\BibitemShut {NoStop}%
\bibitem [{\citenamefont {Beck}\ \emph {et~al.}(2000)\citenamefont {Beck},
  \citenamefont {J\"{a}ckle}, \citenamefont {Worth},\ and\ \citenamefont
  {Meyer}}]{Beck20001}%
  \BibitemOpen
  \bibfield  {author} {\bibinfo {author} {\bibfnamefont {M.}~\bibnamefont
  {Beck}}, \bibinfo {author} {\bibfnamefont {A.}~\bibnamefont {J\"{a}ckle}},
  \bibinfo {author} {\bibfnamefont {G.}~\bibnamefont {Worth}}, \ and\ \bibinfo
  {author} {\bibfnamefont {H.-D.}\ \bibnamefont {Meyer}},\ }\href {\doibase
  10.1016/S0370-1573(99)00047-2} {\bibfield  {journal} {\bibinfo  {journal}
  {Phys. Rep.}\ }\textbf {\bibinfo {volume} {324}},\ \bibinfo {pages} {1 }
  (\bibinfo {year} {2000})}\BibitemShut {NoStop}%
\bibitem [{\citenamefont {Broeckhove}\ \emph {et~al.}(1988)\citenamefont
  {Broeckhove}, \citenamefont {Lathouwers}, \citenamefont {Kesteloot},\ and\
  \citenamefont {Leuven}}]{Broeckhove:1988}%
  \BibitemOpen
  \bibfield  {author} {\bibinfo {author} {\bibfnamefont {J.}~\bibnamefont
  {Broeckhove}}, \bibinfo {author} {\bibfnamefont {L.}~\bibnamefont
  {Lathouwers}}, \bibinfo {author} {\bibfnamefont {E.}~\bibnamefont
  {Kesteloot}}, \ and\ \bibinfo {author} {\bibfnamefont {P.~V.}\ \bibnamefont
  {Leuven}},\ }\href {\doibase 10.1016/0009-2614(88)80380-4} {\bibfield
  {journal} {\bibinfo  {journal} {Chem. Phys. Lett.}\ }\textbf {\bibinfo
  {volume} {149}},\ \bibinfo {pages} {547 } (\bibinfo {year}
  {1988})}\BibitemShut {NoStop}%
\bibitem [{\citenamefont {Mizel}(2004)}]{Mizel:2004}%
  \BibitemOpen
  \bibfield  {author} {\bibinfo {author} {\bibfnamefont {A.}~\bibnamefont
  {Mizel}},\ }\href {\doibase 10.1103/PhysRevA.70.012304} {\bibfield  {journal}
  {\bibinfo  {journal} {Phys. Rev. A}\ }\textbf {\bibinfo {volume} {70}},\
  \bibinfo {pages} {012304} (\bibinfo {year} {2004})}\BibitemShut {NoStop}%
\bibitem [{\citenamefont {Kuprov}\ \emph {et~al.}(2007)\citenamefont {Kuprov},
  \citenamefont {Wagner-Rundell},\ and\ \citenamefont {Hore}}]{Kuprov:2007}%
  \BibitemOpen
  \bibfield  {author} {\bibinfo {author} {\bibfnamefont {I.}~\bibnamefont
  {Kuprov}}, \bibinfo {author} {\bibfnamefont {N.}~\bibnamefont
  {Wagner-Rundell}}, \ and\ \bibinfo {author} {\bibfnamefont {P.}~\bibnamefont
  {Hore}},\ }\href {\doibase 10.1016/j.jmr.2007.09.014} {\bibfield  {journal}
  {\bibinfo  {journal} {J. Magn. Reson.}\ }\textbf {\bibinfo {volume} {189}},\
  \bibinfo {pages} {241 } (\bibinfo {year} {2007})}\BibitemShut {NoStop}%
\bibitem [{\citenamefont {Hogben}\ \emph {et~al.}(2010)\citenamefont {Hogben},
  \citenamefont {Hore},\ and\ \citenamefont {Kuprov}}]{Hogben:2010}%
  \BibitemOpen
  \bibfield  {author} {\bibinfo {author} {\bibfnamefont {H.~J.}\ \bibnamefont
  {Hogben}}, \bibinfo {author} {\bibfnamefont {P.~J.}\ \bibnamefont {Hore}}, \
  and\ \bibinfo {author} {\bibfnamefont {I.}~\bibnamefont {Kuprov}},\ }\href
  {\doibase 10.1063/1.3398146} {\bibfield  {journal} {\bibinfo  {journal} {J.
  Chem. Phys.}\ }\textbf {\bibinfo {volume} {132}},\ \bibinfo {eid} {174101}
  (\bibinfo {year} {2010})}\BibitemShut {NoStop}%
\bibitem [{\citenamefont {Luban}\ \emph {et~al.}(2002)\citenamefont {Luban},
  \citenamefont {Borsa}, \citenamefont {Bud'ko}, \citenamefont {Canfield},
  \citenamefont {Jun}, \citenamefont {Jung}, \citenamefont {K\"ogerler},
  \citenamefont {Mentrup}, \citenamefont {M\"uller}, \citenamefont {Modler},
  \citenamefont {Procissi}, \citenamefont {Suh},\ and\ \citenamefont
  {Torikachvili}}]{Luban:2002}%
  \BibitemOpen
  \bibfield  {author} {\bibinfo {author} {\bibfnamefont {M.}~\bibnamefont
  {Luban}}, \bibinfo {author} {\bibfnamefont {F.}~\bibnamefont {Borsa}},
  \bibinfo {author} {\bibfnamefont {S.}~\bibnamefont {Bud'ko}}, \bibinfo
  {author} {\bibfnamefont {P.}~\bibnamefont {Canfield}}, \bibinfo {author}
  {\bibfnamefont {S.}~\bibnamefont {Jun}}, \bibinfo {author} {\bibfnamefont
  {J.~K.}\ \bibnamefont {Jung}}, \bibinfo {author} {\bibfnamefont
  {P.}~\bibnamefont {K\"ogerler}}, \bibinfo {author} {\bibfnamefont
  {D.}~\bibnamefont {Mentrup}}, \bibinfo {author} {\bibfnamefont
  {A.}~\bibnamefont {M\"uller}}, \bibinfo {author} {\bibfnamefont
  {R.}~\bibnamefont {Modler}}, \bibinfo {author} {\bibfnamefont
  {D.}~\bibnamefont {Procissi}}, \bibinfo {author} {\bibfnamefont {B.~J.}\
  \bibnamefont {Suh}}, \ and\ \bibinfo {author} {\bibfnamefont
  {M.}~\bibnamefont {Torikachvili}},\ }\href {\doibase
  10.1103/PhysRevB.66.054407} {\bibfield  {journal} {\bibinfo  {journal} {Phys.
  Rev. B}\ }\textbf {\bibinfo {volume} {66}},\ \bibinfo {pages} {054407}
  (\bibinfo {year} {2002})}\BibitemShut {NoStop}%
\bibitem [{\citenamefont {Habershon}(2012)}]{Habershon:2012nx}%
  \BibitemOpen
  \bibfield  {author} {\bibinfo {author} {\bibfnamefont {S.}~\bibnamefont
  {Habershon}},\ }\href {\doibase 10.1063/1.3681167} {\bibfield  {journal}
  {\bibinfo  {journal} {J. Chem. Phys.}\ }\textbf {\bibinfo {volume} {136}},\
  \bibinfo {eid} {054109} (\bibinfo {year} {2012})}\BibitemShut {NoStop}%
\bibitem [{\citenamefont {Castro}\ \emph {et~al.}(2004)\citenamefont {Castro},
  \citenamefont {Marques},\ and\ \citenamefont {Rubio}}]{Castro:2004}%
  \BibitemOpen
  \bibfield  {author} {\bibinfo {author} {\bibfnamefont {A.}~\bibnamefont
  {Castro}}, \bibinfo {author} {\bibfnamefont {M.~A.~L.}\ \bibnamefont
  {Marques}}, \ and\ \bibinfo {author} {\bibfnamefont {A.}~\bibnamefont
  {Rubio}},\ }\href {\doibase 10.1063/1.1774980} {\bibfield  {journal}
  {\bibinfo  {journal} {J. Chem. Phys.}\ }\textbf {\bibinfo {volume} {121}},\
  \bibinfo {pages} {3425} (\bibinfo {year} {2004})}\BibitemShut {NoStop}%
\bibitem [{\citenamefont {Smith}\ \emph {et~al.}(2004)\citenamefont {Smith},
  \citenamefont {Bjorstad},\ and\ \citenamefont {Gropp}}]{Smith:2004}%
  \BibitemOpen
  \bibfield  {author} {\bibinfo {author} {\bibfnamefont {B.}~\bibnamefont
  {Smith}}, \bibinfo {author} {\bibfnamefont {P.}~\bibnamefont {Bjorstad}}, \
  and\ \bibinfo {author} {\bibfnamefont {W.}~\bibnamefont {Gropp}},\
  }\href@noop {} {\emph {\bibinfo {title} {Domain decomposition}}}\ (\bibinfo
  {publisher} {Cambridge University Press},\ \bibinfo {year}
  {2004})\BibitemShut {NoStop}%
\bibitem [{\citenamefont {Lions}\ \emph {et~al.}(2001)\citenamefont {Lions},
  \citenamefont {Maday},\ and\ \citenamefont {Turinici}}]{Lions:2001}%
  \BibitemOpen
  \bibfield  {author} {\bibinfo {author} {\bibfnamefont {J.}~\bibnamefont
  {Lions}}, \bibinfo {author} {\bibfnamefont {Y.}~\bibnamefont {Maday}}, \ and\
  \bibinfo {author} {\bibfnamefont {G.}~\bibnamefont {Turinici}},\ }\href
  {\doibase doi:10.1016/S0764-4442(00)01793-6} {\bibfield  {journal} {\bibinfo
  {journal} {Comptes Rendus de l'Academie des Sciences Series I Mathematics}\
  }\textbf {\bibinfo {volume} {332}},\ \bibinfo {pages} {661} (\bibinfo {year}
  {2001})}\BibitemShut {NoStop}%
\bibitem [{\citenamefont {Baffico}\ \emph {et~al.}(2002)\citenamefont
  {Baffico}, \citenamefont {Bernard}, \citenamefont {Maday}, \citenamefont
  {Turinici},\ and\ \citenamefont {Z\'erah}}]{Baffico:2002}%
  \BibitemOpen
  \bibfield  {author} {\bibinfo {author} {\bibfnamefont {L.}~\bibnamefont
  {Baffico}}, \bibinfo {author} {\bibfnamefont {S.}~\bibnamefont {Bernard}},
  \bibinfo {author} {\bibfnamefont {Y.}~\bibnamefont {Maday}}, \bibinfo
  {author} {\bibfnamefont {G.}~\bibnamefont {Turinici}}, \ and\ \bibinfo
  {author} {\bibfnamefont {G.}~\bibnamefont {Z\'erah}},\ }\href {\doibase
  10.1103/PhysRevE.66.057701} {\bibfield  {journal} {\bibinfo  {journal} {Phys.
  Rev. E}\ }\textbf {\bibinfo {volume} {66}},\ \bibinfo {pages} {057701}
  (\bibinfo {year} {2002})}\BibitemShut {NoStop}%
\bibitem [{\citenamefont {Gander}\ and\ \citenamefont
  {Vandewalle}(2007)}]{Gander:2007}%
  \BibitemOpen
  \bibfield  {author} {\bibinfo {author} {\bibfnamefont {M.~J.}\ \bibnamefont
  {Gander}}\ and\ \bibinfo {author} {\bibfnamefont {S.}~\bibnamefont
  {Vandewalle}},\ }\href@noop {} {\bibfield  {journal} {\bibinfo  {journal}
  {SIAM J. Sci. Comp.}\ }\textbf {\bibinfo {volume} {29}},\ \bibinfo {pages}
  {556} (\bibinfo {year} {2007})}\BibitemShut {NoStop}%
\bibitem [{\citenamefont {Gander}(2008)}]{Gander:2008}%
  \BibitemOpen
  \bibfield  {author} {\bibinfo {author} {\bibfnamefont {M.~J.}\ \bibnamefont
  {Gander}},\ }\href@noop {} {\bibfield  {journal} {\bibinfo  {journal}
  {Bolet{\'\i}n SEMA}\ } (\bibinfo {year} {2008})}\BibitemShut {NoStop}%
\bibitem [{\citenamefont {Hestenes}\ and\ \citenamefont
  {Stiefel}(1952)}]{Hestenes:1952}%
  \BibitemOpen
  \bibfield  {author} {\bibinfo {author} {\bibfnamefont {M.~R.}\ \bibnamefont
  {Hestenes}}\ and\ \bibinfo {author} {\bibfnamefont {E.}~\bibnamefont
  {Stiefel}},\ }\href@noop {} {\enquote {\bibinfo {title} {Methods of conjugate
  gradients for solving linear systems},}\ } (\bibinfo {year}
  {1952})\BibitemShut {NoStop}%
\bibitem [{\citenamefont {Faber}\ and\ \citenamefont
  {Manteuffel}(1984)}]{Faber:1984}%
  \BibitemOpen
  \bibfield  {author} {\bibinfo {author} {\bibfnamefont {V.}~\bibnamefont
  {Faber}}\ and\ \bibinfo {author} {\bibfnamefont {T.}~\bibnamefont
  {Manteuffel}},\ }\href@noop {} {\bibfield  {journal} {\bibinfo  {journal}
  {SIAM J. Numer. Anal.}\ }\textbf {\bibinfo {volume} {21}},\ \bibinfo {pages}
  {352} (\bibinfo {year} {1984})}\BibitemShut {NoStop}%
\bibitem [{\citenamefont {Eisenstat}(1981)}]{Eisenstat:1981}%
  \BibitemOpen
  \bibfield  {author} {\bibinfo {author} {\bibfnamefont {S.~C.}\ \bibnamefont
  {Eisenstat}},\ }\href@noop {} {\bibfield  {journal} {\bibinfo  {journal}
  {SIAM J. Sci. Stat. Comp.}\ }\textbf {\bibinfo {volume} {2}},\ \bibinfo
  {pages} {1} (\bibinfo {year} {1981})}\BibitemShut {NoStop}%
\bibitem [{\citenamefont {Makri}(1993)}]{Makri:1993}%
  \BibitemOpen
  \bibfield  {author} {\bibinfo {author} {\bibfnamefont {N.}~\bibnamefont
  {Makri}},\ }\href@noop {} {\bibfield  {journal} {\bibinfo  {journal} {J.
  Phys. Chem.}\ }\textbf {\bibinfo {volume} {97}},\ \bibinfo {pages} {2417}
  (\bibinfo {year} {1993})}\BibitemShut {NoStop}%
\bibitem [{\citenamefont {Feit}\ \emph {et~al.}(1982)\citenamefont {Feit},
  \citenamefont {Jr.},\ and\ \citenamefont {Steiger}}]{Feit:1982}%
  \BibitemOpen
  \bibfield  {author} {\bibinfo {author} {\bibfnamefont {M.}~\bibnamefont
  {Feit}}, \bibinfo {author} {\bibfnamefont {J.~F.}\ \bibnamefont {Jr.}}, \
  and\ \bibinfo {author} {\bibfnamefont {A.}~\bibnamefont {Steiger}},\
  }\href@noop {} {\bibfield  {journal} {\bibinfo  {journal} {J. Comput. Phys.}\
  }\textbf {\bibinfo {volume} {47}},\ \bibinfo {pages} {412} (\bibinfo {year}
  {1982})}\BibitemShut {NoStop}%
\bibitem [{\citenamefont {Westermann}\ and\ \citenamefont
  {Manthe}(2012)}]{Westermann:2012dq}%
  \BibitemOpen
  \bibfield  {author} {\bibinfo {author} {\bibfnamefont {T.}~\bibnamefont
  {Westermann}}\ and\ \bibinfo {author} {\bibfnamefont {U.}~\bibnamefont
  {Manthe}},\ }\href {\doibase 10.1063/1.4733676} {\bibfield  {journal}
  {\bibinfo  {journal} {J. Chem. Phys.}\ }\textbf {\bibinfo {volume} {137}},\
  \bibinfo {eid} {22A509} (\bibinfo {year} {2012})}\BibitemShut {NoStop}%
\bibitem [{\citenamefont {Helgaker}\ \emph {et~al.}(2002)\citenamefont
  {Helgaker}, \citenamefont {Jorgensen},\ and\ \citenamefont
  {Olsen}}]{Helgaker:2002}%
  \BibitemOpen
  \bibfield  {author} {\bibinfo {author} {\bibfnamefont {T.}~\bibnamefont
  {Helgaker}}, \bibinfo {author} {\bibfnamefont {P.}~\bibnamefont {Jorgensen}},
  \ and\ \bibinfo {author} {\bibfnamefont {J.}~\bibnamefont {Olsen}},\
  }\href@noop {} {\emph {\bibinfo {title} {Molecular Electronic Structure
  Theory}}}\ (\bibinfo  {publisher} {Wiley, Sussex},\ \bibinfo {year}
  {2002})\BibitemShut {NoStop}%
\bibitem [{\citenamefont {Kvaal}(2012)}]{Kvaal:2012cr}%
  \BibitemOpen
  \bibfield  {author} {\bibinfo {author} {\bibfnamefont {S.}~\bibnamefont
  {Kvaal}},\ }\href {\doibase 10.1063/1.4718427} {\bibfield  {journal}
  {\bibinfo  {journal} {J. Chem. Phys.}\ }\textbf {\bibinfo {volume} {136}},\
  \bibinfo {eid} {194109} (\bibinfo {year} {2012})}\BibitemShut {NoStop}%
\bibitem [{\citenamefont {Sambe}(1973)}]{Sambe:1973}%
  \BibitemOpen
  \bibfield  {author} {\bibinfo {author} {\bibfnamefont {H.}~\bibnamefont
  {Sambe}},\ }\href {\doibase 10.1103/PhysRevA.7.2203} {\bibfield  {journal}
  {\bibinfo  {journal} {Phys. Rev. A}\ }\textbf {\bibinfo {volume} {7}},\
  \bibinfo {pages} {2203} (\bibinfo {year} {1973})}\BibitemShut {NoStop}%
\bibitem [{\citenamefont {Howland}(1974)}]{Howland:1974}%
  \BibitemOpen
  \bibfield  {author} {\bibinfo {author} {\bibfnamefont {J.~S.}\ \bibnamefont
  {Howland}},\ }\href {http://dx.doi.org/10.1007/BF01351346} {\bibfield
  {journal} {\bibinfo  {journal} {Mathematische Annalen}\ }\textbf {\bibinfo
  {volume} {207}},\ \bibinfo {pages} {315} (\bibinfo {year}
  {1974})}\BibitemShut {NoStop}%
\bibitem [{\citenamefont {Shalashilin}\ and\ \citenamefont
  {Burghardt}(2008)}]{Shalashilin:2008}%
  \BibitemOpen
  \bibfield  {author} {\bibinfo {author} {\bibfnamefont {D.~V.}\ \bibnamefont
  {Shalashilin}}\ and\ \bibinfo {author} {\bibfnamefont {I.}~\bibnamefont
  {Burghardt}},\ }\href {\doibase 10.1063/1.2969101} {\bibfield  {journal}
  {\bibinfo  {journal} {J. Chem. Phys.}\ }\textbf {\bibinfo {volume} {129}},\
  \bibinfo {eid} {084104} (\bibinfo {year} {2008})}\BibitemShut {NoStop}%
\end{thebibliography}%
\end{document}